\documentclass[lettersize,journal]{IEEEtran}
\usepackage{cite}
\usepackage{amsmath,amssymb,amsfonts}
\usepackage{graphicx}
\usepackage{textcomp}
\usepackage{xcolor}
\usepackage{algorithm}
\usepackage{algorithmicx}
\usepackage{algpseudocode}
\usepackage{todonotes}
 
\usepackage{amsmath,amssymb,amsfonts}
\usepackage{ulem}
\usepackage{booktabs}
\usepackage{float}
\usepackage{multirow}
\usepackage{subfigure}
\usepackage{listings}
\usepackage{pifont}
\usepackage{enumerate}
\usepackage{hyperref} 
\usepackage{graphicx}
\usepackage{pifont}
\usepackage{CJKutf8}
\usepackage{url}

\usepackage{fancyhdr}
\fancypagestyle{preContent}{
    \fancyhead{}
    
    \fancyfoot[C]{\thepage}
}
\graphicspath{{Figures/}}

\hypersetup{
    colorlinks = true,
    linkcolor=blue,
    filecolor=blue,      
    urlcolor=blue,
    citecolor=cyan,
}

\begin{document}

\title{PWDFT-SW: Extending the Limit of Plane-Wave DFT Calculations to 16K Atoms on the New Sunway Supercomputer}

\author{\IEEEauthorblockN{
Qingcai Jiang\IEEEauthorrefmark{2}, Zhenwei Cao\IEEEauthorrefmark{2}, Junshi Chen, Xinming Qin, 
\\Wei Hu\IEEEauthorrefmark{1}, Hong An\IEEEauthorrefmark{1} and Jinlong Yang}
    \IEEEcompsocitemizethanks{
        % \IEEEcompsocthanksitem Q. Jiang, Z. Cao, J. Chen, X. Qin, W. Hu, H. An and J. Yang are with the University of Science and Technology of China. 
        \IEEEcompsocthanksitem Qingcai Jiang is with the University of Science and Technology of China, Hefei, Anhui 230026, China. E-mail: jqc@mail.ustc.edu.cn.
        \IEEEcompsocthanksitem Zhenwei Cao is with the University of Science and Technology of China, Hefei, Anhui 230026, China and Kuaishou Technology, Beijing 100085, China. E-mail: caozhenwei@kuaishou.com
        \IEEEcompsocthanksitem Junshi Chen and Hong An are with the University of Science and Technology of China, Hefei, Anhui 230026, China and Laoshan Laboratory, Qingdao, Shandong 266237, China. 
        \IEEEcompsocthanksitem Xinming Qin, Wei Hu and Jinlong Yang are with State Key Laboratory of Precision and Intelligent Chemistry, University of Science and Technology of China, Hefei, Anhui 230026, China.
        \IEEEcompsocthanksitem \IEEEauthorrefmark{2} Equal contributions.
        \IEEEcompsocthanksitem \IEEEauthorrefmark{1} Corresponding authors. E-mail: \{whuustc, han\}@ustc.edu.cn.
    }
    % \thanks{
    %     \IEEEauthorrefmark{2} Equal contributions.\\
    %     \IEEEauthorrefmark{1} Corresponding authors. E-mail: \{whuustc, han, jlyang\}@ustc.edu.cn.\\}
}
 
\markboth{Journal of \LaTeX\ Class Files,~Vol.~14, No.~8, August~2021}%
{Shell \MakeLowercase{\textit{et al.}}: A Sample Article Using IEEEtran.cls for IEEE Journals}


\maketitle

\begin{abstract}
First-principles density functional theory (DFT) with plane wave (PW) basis set is the most widely used method in quantum mechanical material simulations due to its advantages in accuracy and universality. 
However, a perceived drawback of PW-based DFT calculations is their substantial computational cost and memory usage, which currently limits their ability to simulate large-scale complex systems containing thousands of atoms. 
This situation is exacerbated in the new Sunway supercomputer, where each process is limited to a mere 16 GB of memory. 
Herein, we present a novel parallel implementation of plane wave density functional theory on the new Sunway supercomputer (PWDFT-SW).
PWDFT-SW fully extracts the benefits of Sunway supercomputer by extensively refactoring and calibrating our algorithms to align with the system characteristics of the Sunway system.
Through extensive numerical experiments, we demonstrate that our methods can substantially decrease both computational costs and memory usage.
Our optimizations translate to a speedup of 64.8x for a physical system containing 4,096 silicon atoms, enabling us to push the limit of PW-based DFT calculations to large-scale systems containing 16,384 carbon atoms.

% This document describes the most common article elements and how to use the IEEEtran class with \LaTeX \ to produce files that are suitable for submission to the IEEE.  IEEEtran can produce conference, journal, and technical note (correspondence) papers with a suitable choice of class options. 
\end{abstract}

\begin{IEEEkeywords}
Quantum mechanics, Plane-wave density functional theory, New Sunway supercomputer, Parallel implementation.
% Article submission, IEEE, IEEEtran, journal, \LaTeX, paper, template, typesetting.
\end{IEEEkeywords}

\section{Introduction}

\IEEEPARstart{A}{mong} the major applications in high-performance computing (HPC), first-principles density functional theory (DFT) calculations hold significant importance for quantum mechanical simulations of materials~\cite{carter2008challenges}. DFT enables a direct connection between atomic structures and their associated physical properties, underpinning research in numerous scientific and engineering fields~\cite{hohenberg1964inhomogeneous}. Given the critical role of DFT, extensive efforts have been made to develop efficient and accurate numerical approaches to solve the Kohn-Sham equations~\cite{nakata2020large,shang2021extreme,jia2013fast,das2019fast,hasegawa2011first,wang2008linearly,kuhne2020cp2k,banerjee2018two}.
Several numerical discretization methods exist for representing wavefunctions in DFT calculations, including numerical atomic orbital (NAO) basis sets~\cite{hehre1969self,jensen2002polarization,hutter2014wiley} and Gaussian-type orbital (GTO) basis sets~\cite{frisch1984self}. 
However, these localized orbital-based approaches are typically not systematically convergent for generic material systems. 
In contrast, the plane wave (PW) basis set~\cite{gygi2006large, giannozzi2009quantum, hu2017adaptively} is widely favored due to its systematic convergence, accuracy, completeness, and orthogonality, making it particularly suitable even for strongly correlated systems. 
Consequently, PW-based DFT codes, such as the widely used commercial software VASP~\cite{kresse1996efficient,hafner2008ab} and the open-source package Quantum Espresso~\cite{giannozzi2009quantum,giannozzi2017advanced}, have gained extensive popularity in computational material science.

Recently, real-space discretization methods, such as SPARC~\cite{xu2021sparc} and DFT-FE~\cite{motamarri2020dft}, have demonstrated significant advantages in HPC environments due to their excellent parallel scalability and reduced communication overhead. 
These real-space methods avoid the global communication bottlenecks associated with the Fast Fourier Transform (FFT) operations required by PW methods. 
Nevertheless, the maturity, established software ecosystem, and robust convergence properties of PW-based methods continue to make them highly attractive and widely adopted in current research. 
In this work, we focus on the plane-wave-based PWDFT framework, aiming to exploit these advantages while enhancing computational efficiency and scalability.

% ---- Delete for the Euro-Par submission ---- 
Despite the popularity of plane wave basis set, the ultra high computation and memory cost of plane wave based DFT software remain a main stumbling block to the mainstream adoption of this highly accurate method.
Conventional DFT calculations with plane wave basis set take $\mathcal{O}(N^{3})$ floating point operations and cost $\mathcal{O}(N^2)$ memory footprint regarding the number of electrons in the studied physical system $N$.
For example, the number of basis $N_\textrm{PW}$ in PW-based DFT calculations for a moderate-scale physical system with less than 5K atoms is approximately above $10^8$, which takes at least $10^{16}$ bytes of memory footprint and $10^{24}$ floating point operations. 
Therefore, it is hard for state-of-the-art PW-based DFT software to simulate large systems containing thousands of atoms even on modern supercomputers~\cite{barnes2017improved,zhao2017performance}. % ~\cite{barnes2017improved,zhao2017performance}
As a result, exploring the physical properties of systems with thousands of atoms within PW-based DFT framework remains challenging.

% There are typically two types of method to push the limit of DFT calculations: (1) developing reduced order basis and (2) leveraging the the tremendous performance offered by modern supercomputers. 
% There are a number of previous works aimed to develop small-size localized basis sets, such as numerical atomic orbitals (NAO) basis sets~\cite{hehre1969self,jensen2002polarization,hutter2014wiley} and Gaussian-type orbital (GTO) basis sets~\cite{frisch1984self}. 
% However, a perceived drawback of these type of basis sets is that they are not systematically convergent for generic materials systems. 

Hence, a lot of research focuses on expanding the system size of PW-based DFT calculations on modern high-performance computing facilities~\cite{jia2019parallel, igram2018large}. 
The new Sunway supercomputer is gaining popularity due to its ultra-high floating point performance (1.5 EFLOPS in total of the theoretical peak performance), which provides a promising way to push the limits of PW-based DFT calculations if we can fully leverage this supercomputer. 
However, there has been very little work on massively parallel PW-based DFT calculations, particularly on the new Sunway supercomputer, due to several challenges:
\begin{itemize}
\item Very limited 16 GB memory available for each process in Sunway supercomputer,
\item  Massive all-to-all memory accessing and communications in PW-based DFT software
\item  The complex programming scheme in Sunway supercomputer.
\end{itemize}

Plane Wave Density Functional Theory (PWDFT)~\cite{hu2017adaptively} is one of the most scalable PW-based parallel DFT calculation frameworks to date. Based on our comprehensive characterization of PWDFT on the new Sunway supercomputer, we make two key observations:
(1) A significant portion of the memory cost (e.g., temporary buffers and redundant pseudopotential information) in traditional PW-based DFT calculations can be alleviated with negligible communication cost.
(2) Excessive small messages in PW-based DFT calculations incur heavy penalties due to traffic overload in communications and memory access.

Based on our observations, we revamp PWDFT on the new Sunway supercomputer in the following aspects:
(1) We significantly reduce memory usage by efficiently refactoring the calculation and application of pseudopotentials with negligible communication cost.
(2) We pack elements and use in-place data transformation to eliminate many temporary buffers and frequent small data transfers, leading to up to a 14x speedup.
(3) We establish a multistage Allreduce operator, which leverages the characteristics of both hardware and the underlying communication library, enabling more comprehensive exploitation of network bandwidth performance.
(4) We propose granularity-aware parallel mechanisms that consider the computation and communication patterns of different steps in PWDFT to select the most suitable parallel scale, further boosting the execution of PWDFT.
(5) We establish a SWUC-assisted parallel programming framework to reduce the cumbersome programming handwork on the new Sunway platform.
Additionally, we make some minor contributions to PWDFT, such as optimizing the data structure to make data access for CPEs more friendly and implementing a C++ library in PWDFT to monitor memory occupation at runtime.

The main contributions of this work can be summarized as follows:
\begin{itemize}
\item To address the key challenges of PW-based DFT code on the new Sunway supercomputer, we propose several optimization methods to leverage the limited 16 GB memory capacity of each process for massively parallel implementation of PWDFT code. Additionally, we devise several mechanisms to fully accelerate PWDFT.
\item Through extensive numerical experiments, we demonstrate that our methods can achieve high scalability in both strong scaling and weak scaling. Furthermore, we achieve a speedup of 64.8 for a physical system containing 4,096 silicon atoms compared to the directly transported version, which is a very convincing result in PW-based DFT codes.
\item We show that with our algorithms, along with parallel implementations and optimizations, we can study the graphene system with 16,384 atoms. This result greatly outperforms state-of-the-art PW-based DFT codes in terms of the physical system scale.
\end{itemize}

Most of the optimization methods and tools proposed in this work could extend to other PW-based DFT codes with the same computational characteristics, not only on the Sunway platform but also on other HPC facilities.

\section{Background}
\subsection{Background of PWDFT}
PWDFT aims to solve an eigenvalue problem of this form:
\begin{equation}
    H X=X \Lambda,
\end{equation}
where H denotes the PW Hamiltonian, X presents the wavefunctions discretized with plane waves, and the eigenvalues of this problem $\lambda$ are energies of the studied system. 

\begin{figure}[htbp]
    \centering
    \includegraphics[width=\linewidth]{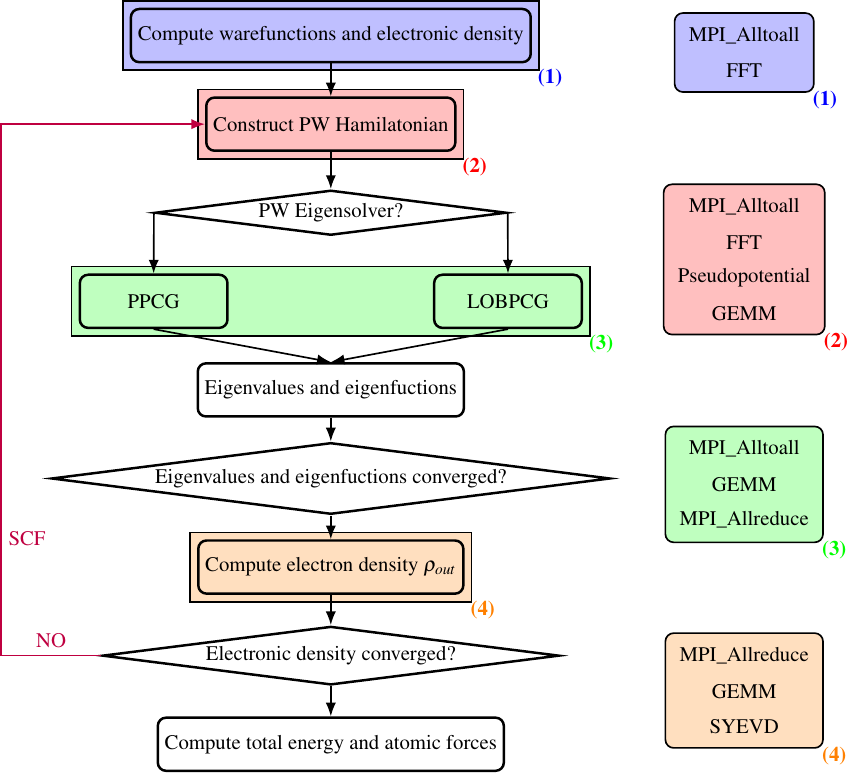}
    \caption{Basic flowchart and 4 time-consuming steps with their corresponding numerical kernels (in the same background color) of PWDFT.}
    \label{fig:pwdft}
\end{figure}

Similar to many other DFT codes, PWDFT solves the eigenvalue problem in a self-consistent field (SCF) method. 
The whole SCF iterations in PWDFT contain two parts: (1) the inner eigenvalue problem of PW Hamiltonian and (2) SCF for checking the convergence of energy~\cite{1992Iterative}, as shown in Figure~\ref{fig:pwdft}.
\textbf{SCF} part, the Kohn-Sham equation $H\Psi_i=\varepsilon_i \Psi_i$ is solved to obtain the improvement of the wavefunction $\Psi_i$ for a given particle Hamiltonian $H_{i}$, and the wavefunction is expanded based on plane wave basis set. 
The SCF terminates when the local potential $V_{Hxc}$ agrees to the input wavefunctions. 
\textbf{The inner eigenvalue problem} is mainly responsible for diagonalizing the PW Hamiltonian by an iterative conjugate gradient eigensolver, PPCG or LOBPCG. 
The inner eigenvalue problem begins when the Hamiltonian matrix is constructed by wavefunctions, which aim to solve eigenproblems and acquire eigenvalues and eigenfunctions. 
Then the charge density $\rho\left(r\right)$ is calculated by eigenfunctions according to $\rho\left(r\right)=\sum_{j=1}^{N_e}\left|\psi_j\right|^2$, where $\psi_j$ is the j-th eigenfunction and $N_e$ represents the total valence charge number. 
Then exchange-correlation potential $V_{Hxc}$ is calculated to be compared with the input for consistency. 
We present the major numerical kernels in each step on PWDFT on the right of Figure~\ref{fig:pwdft} to enhance the readability.

In theory, wave functions are constructed using an infinite plane wave basis. However, $E_{cut}$ is set to include waves with energy less than a certain input value. This cutoff energy determines the grid number in real space $N_r$ according to $\left(N_r\right)_i=\sqrt{2 E_{cut}} L_i / \pi$, where $L_i$ is the length of the supercell in different dimensions (x, y, and z).
In our experiments, the default value of $E_{cut}$ is set to 5 Hartree unless otherwise specified. 
Table~\ref{tab:gridnum} lists the grid number for certain systems in our experiments.

\begin{table}[htbp]
\centering
\caption{Grid Number of Typical Test Cases.}
\label{tab:gridnum}
\begin{tabular}{|c|c|c|c|c|}
\hline
System   & x-axis  & y-axis   & z-axis   & total   \\ \hline
Si$_{4096}$ & 84 & 84  & 84  & 592704  \\ \hline
Si$_{8192}$ & 88 & 88  & 176 & 1362944 \\ \hline
G$_{11520}$ & 16 & 164 & 170 & 446080  \\ \hline
\end{tabular}
\end{table}

\subsection{Parallelization Scheme}
\begin{figure}[htbp]
    \centering
    \includegraphics[width=\linewidth]{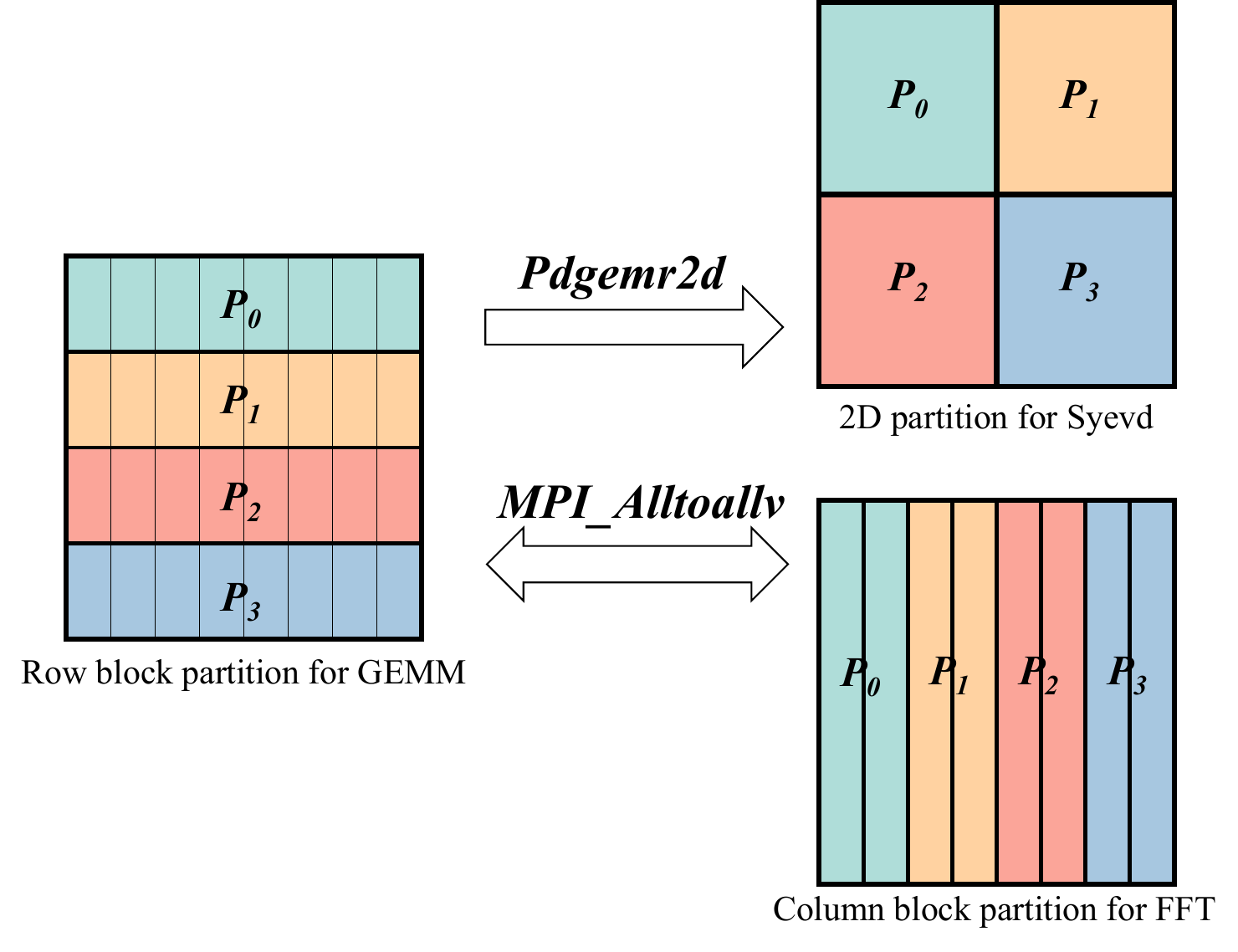}
    \caption{Parallel data and task distribution schemes.}
    \label{fig:alltoall}
\end{figure}

There are three main parallelization schemes as shown in Figure \ref{fig:alltoall} due to different intrinsic physical natures in PWDFT. The parallelization scheme is based on the distribution of the wavefunction matrix. 
\textbf{The first scheme} is the row block partition. This scheme facilitates matrix-matrix multiplications such as $S=\Psi^*(H \Psi)$ when $H \Psi$ is obtained.
\textbf{The second scheme} is the column block partition. In this scheme, each process holds some entire wavefunctions, allowing a single process to perform a fast FFT without any data transformation or communication. This data distribution pattern also simplifies the calculation of \(H\Psi\), where \( H\) is the Hamiltonian operator applied to wavefunctions.
\textbf{The third scheme} is a 2-dimensional data partition used exclusively for Syevd, a ScaLAPACK function to acquire eigenvalues and eigenvectors.
We perform MPI\_Alltoallv and a ScaLAPACK routine Pdgemr2d to transform data patterns among the 3 schemes.

\subsection{Problem Overview}
\label{sec:problem_overview}
In this section, we will elaborate on the three main obstacles that hinder the efficient parallelization of PWDFT on the new Sunway supercomputer.
\subsubsection{\textbf{Limited Memory Size of Each Process}}
\begin{figure}[htbp]
    \centering
    \includegraphics[width=\linewidth]{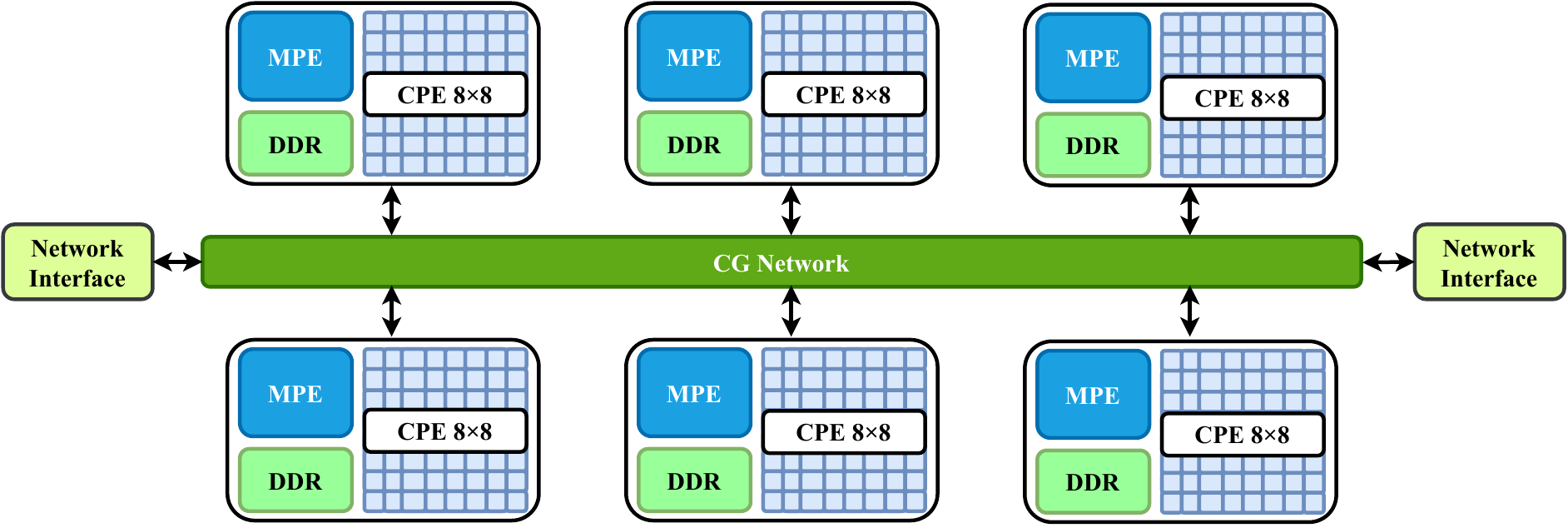}
    \caption{Architecture of SW26010Pro Processor.}
    \label{fig:sunway}
\end{figure}

Figure~\ref{fig:sunway} depicts the architecture of the SW26010Pro processor in the new Sunway supercomputer, comprising six core groups (CGs) each with one management processor element (MPE) and 64 computing processor elements (CPEs). 
Each processor is equipped with 96 GB of memory, evenly allocated at 16 GB per CG. 
The new Sunway supercomputer employs two memory partition schemes: (1) the common scheme, where each core group acts as a separate processor with 16 GB of memory, and (2) the large memory scheme, where six core groups in a process share 96 GB of memory and act as a single process. 
However, numerical libraries that are extensively used in PWDFT like FFTW~\cite{frigo1998fftw}, BLAS~\cite{dongarra1990set}, and LAPACK~\cite{angerson1990lapack} are tailored to the common scheme, implying that each process of PW-based DFT can only use the common memory partition scheme, which allocates a very limited 16 GB memory footprint per process.

% The architecture of the many-core processor SW26010Pro on the new Sunway supercomputer is shown in Figure~\ref{fig:sunway}. It consists of 6 core groups (CGs), in which there is one management processor element (MPE) and $8×8$ computing processor elements (CPEs).
% The SW26010Pro processor has a memory capacity of 96 gigabytes (GB), which is divided equally among six core groups, with each group having 16 GB of memory.
% The new Sunway supercomputer employs two memory partition schemes for running applications: 
% (1) the common scheme, where each core group acts as a separate processor with 16 GB of memory, and 
% (2) the large memory scheme, where six core groups in a process share 96 GB of memory and act as a single process.
% Conventional DFT software relies heavily on numerical libraries such as FFTW~\cite{frigo1998fftw}, BLAS~\cite{dongarra1990set} and LAPACK~\cite{angerson1990lapack} etc. 
% However, these numerical libraries only support the common scheme, which implies that each process of PW-based DFT, distributed on each CG, can only allocate a very limited 16 GB memory footprint.

\subsubsection{\textbf{Massive All-to-all Data Communications}}
\label{sec:massive_comm}
The predecessors of the new Sunway, specifically the Sunway Taihu Light, had a peak memory bandwidth of 136 GB/s, creating a significant bottleneck for various applications. 
In contrast, the new Sunway boasts a bandwidth transfer from global memory to the LDM of CPE around 307 GB/s via direct memory access (DMA). 
The bandwidth among the CPEs within the CPE cluster, via remote memory access (RMA), has a theoretical peak performance of 460 GB/s~\cite{0Enabling}. 
However, data sizes per transfer must exceed 1024 and 256 bytes to achieve 80\% bandwidth utilization for DMA and RMA, respectively.

PW-based DFT calculations require extensive all-to-all data communications due to the intrinsic physical properties of the huge dense Hamiltonian matrix. 
These communications remain costly even on the new Sunway supercomputer, despite specialized hardware and software support. Additionally, the transformation of the parallelization scheme involves substantial small-volume and irregular memory access to ensure data continuity. 
These memory accesses are time-consuming due to their small granularity and inadequate data locality, hindering the full utilization of the cache hierarchy. 
Consequently, heavy all-to-all data communications can severely degrade performance in PW-based DFT calculations.

\subsubsection{\textbf{Complicated Programming Framework}}
The major computing resource of the new Sunway supercomputer is provided by the 64 CPEs of each core group. 
To effectively leverage their significant floating-point performance, programmers usually need to manually orchestrate the CPEs through a fine-grained Athread programming interface.
This interface is an effective but cumbersome thread library provided by the vendor.

\section{Mechanisms}
\label{sec:innovations}
In this section, we introduce the mechanisms we propose to address the aforementioned challenges described in Section~\ref{sec:problem_overview}.
\subsection{Refactor of the pseudopotential calculations}
\label{subsec:pseudo}

Pseudopotentials are commonly used in DFT calculations to approximate the atomic nucleus and inner-shell electrons as ion entities, a practice that is justified given that chemical properties are primarily influenced by the behavior of the outermost valence electrons~\cite{milman2000electronic, yin1982theory}.

In Plane Wave Density Functional Theory (PWDFT) software, each atom is represented by specific pseudopotential information that adjusts each wave function accordingly. 
Originally, each computing process maintained the entire dataset of pseudopotentials. 
While the memory required for a single atom's pseudopotential is not substantial, the total memory consumed by all atoms' pseudopotentials significantly limits PWDFT's ability to model larger systems. 
This limitation is particularly acute on the new Sunway system, which has a restricted memory capacity of 16 GB. 
To address this, we have redesigned the pseudopotential calculations by implementing a distributed storage scheme to reduce the memory bottleneck.
 
\begin{figure*}[htbp]
    \centering
    \includegraphics[width=\linewidth]{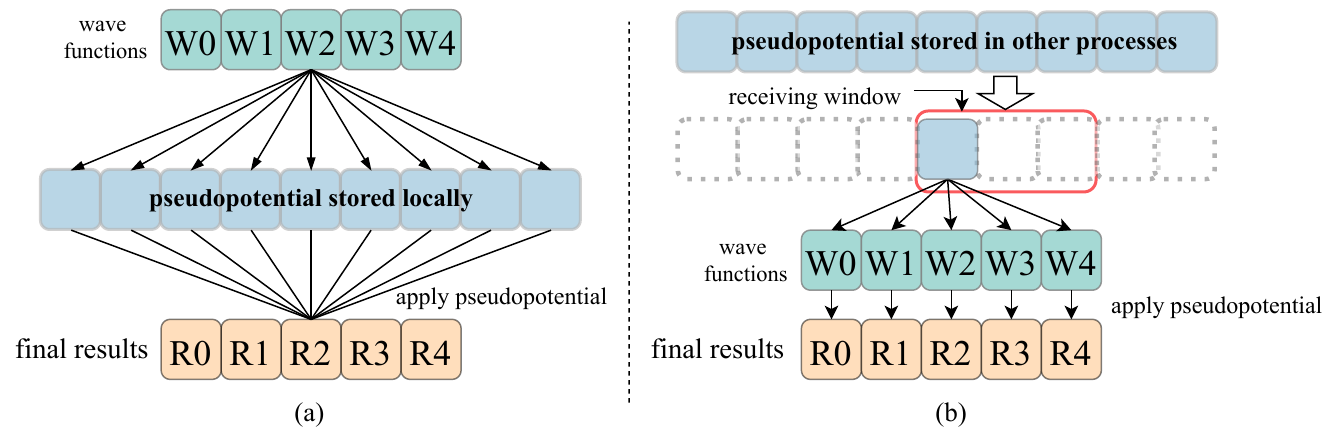}
    \caption{Refactor of pseudopotential process: (a) naive implementation where each process holds the entire pseudopotential, (b) the refactored approach where each process holds a portion of the pseudopotential, in which the dotted, uncolored blocks indicate the absence of pseudopotential in the process.}
    \label{fig:pseudo}
\end{figure*}

The original procedure of applying a pseudopotential to wave functions is illustrated in Figure~\ref{fig:pseudo} (a). These functions are distributed across various processes. 
For each wave function stored locally, the algorithm sequentially processes every atom, adjusting the wave function according to the corresponding pseudopotential. When the pseudopotential is distributed across processes, each process retains only a portion of the overall pseudopotential data. Therefore, to apply the pseudopotential, it has to be retrieved from other processes through a network interface. In the case of a basic algorithm, this necessitates the redundant transmission of the same atomic pseudopotential across distinct wave functions.

Therefore, we apply the pseudopotential of an atom to all local wave functions once obtained, thereby eliminating redundant transmission of the pseudopotential. Additionally, we employ a sliding window algorithm to address significant communication overhead, enabling overlapping computation and communication. This approach, illustrated in Figure~\ref{fig:pseudo} (b), allows a process to receive the upcoming pseudopotential within the sliding window while applying the current pseudopotential to the local wave functions. The communication overhead can be fully overlapped with the flexible window size.

This optimization significantly reduces memory usage. For instance, in a graphene system containing 11,520 atoms, the pseudopotential of a single atom is roughly 0.3 MB when $E_{cut} = 10$. 
In the naive implementation, each process must store the pseudopotential of all atoms, leading to an estimated memory footprint of approximately 3.4 GB. 
This constitutes about 21\% of the entire memory per process. However, once we refactor the pseudopotential process, the memory usage becomes insignificant.

\subsection{Memory Accessing Optimization}
\label{subsec:alltoall}

\subsubsection{\textbf{The mapping and buffer matrix}}
As mentioned, the parallelization scheme that involves both row and column partitioning requires global communication through MPI\_Alltoallv, as illustrated in Figure~\ref{fig:alltoall}. 
However, this global communication via MPI\_Alltoallv in PWDFT to transform data between row and column partitions introduces data transformation to \textbf{make the data contiguous in memory}.

\begin{figure}[htbp]
    \centering
    \includegraphics[width=\linewidth]{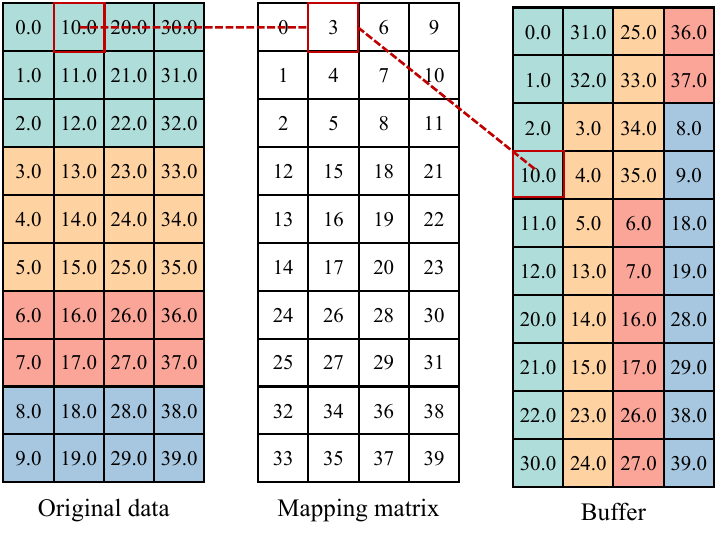}
    \caption{An example of the mapping matrix in the data transformation.}
    \label{fig:sendrecvmap}
\end{figure}

For simplicity, we take the data transformation from column partition to row partition as an example. The data transformation in the naive implementation of PWDFT involves two mapping matrices and buffers, one before and one after the MPI\_Alltoallv operation.
We illustrate the data transformation formula in Figure~\ref{fig:sendrecvmap}, where blocks of the same color are sent to the same processor, and the number in each block of the original data and buffer represents each data element.
For example, the original data matrix in Figure~\ref{fig:sendrecvmap} shows the column partition data in one of the four processes, which is sent to four processes, indicated by different colors. 
In the column-major format, the data block in light turquoise is not contiguous. Thus, the block containing "10.0" in the original data matrix is mapped to the 3rd position (starting from 0) in the buffer matrix in Figure~\ref{fig:sendrecvmap} for MPI\_Alltoallv.

\subsubsection{\textbf{Optimizations}}
\label{subsec:optimization_buffer}

\begin{figure}[htbp]
    \centering
    \includegraphics[width=\linewidth]{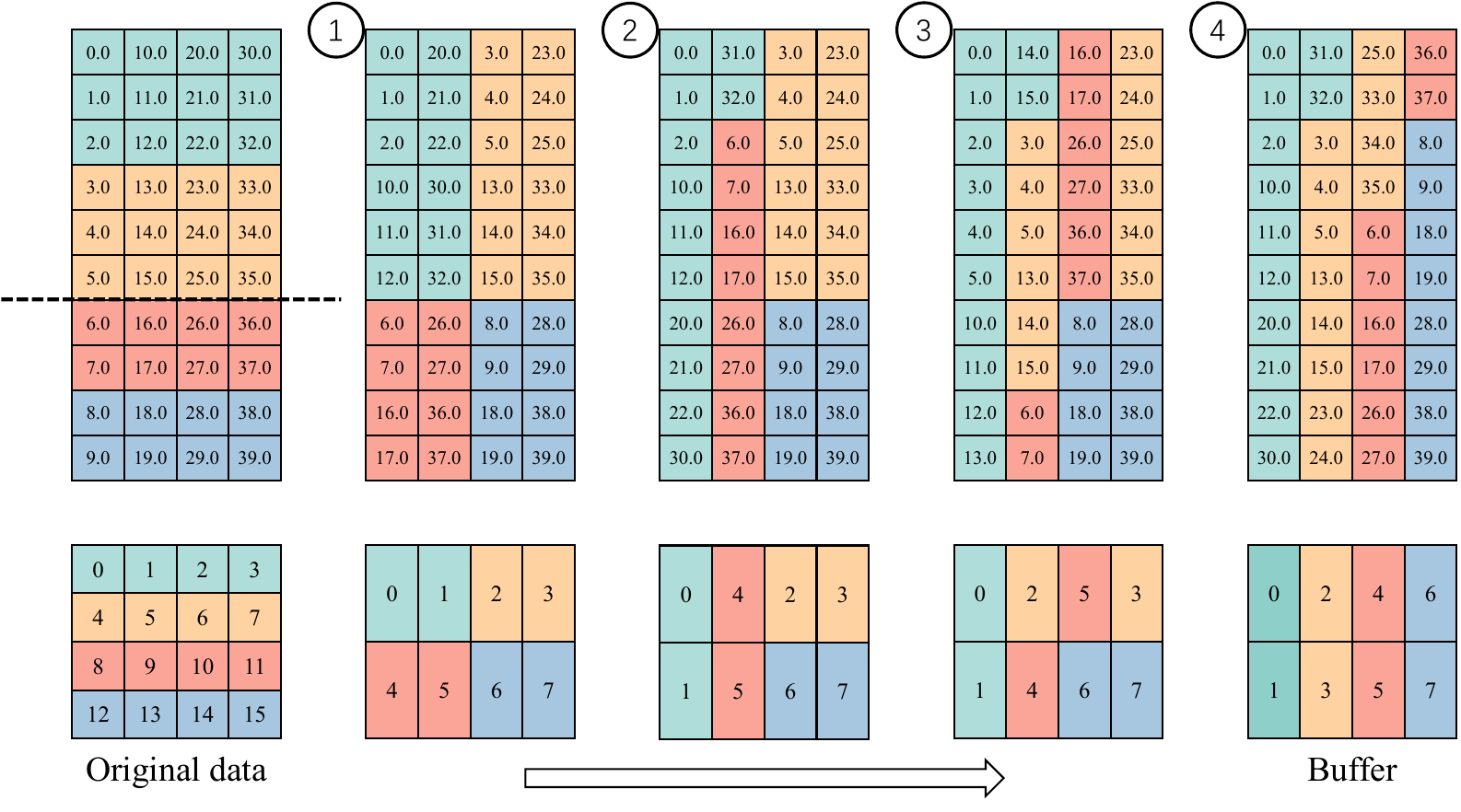}
    \caption{An example of the in-place data transformation. The circled numbers denote the procedures in the transformation.}
    \label{fig:irregularbuf}
\end{figure}

Here, we design an in-place data transformation, as illustrated in Figure~\ref{fig:irregularbuf}, to reduce the heavy memory occupation of the buffer.

We combine the data in the same column of each process into \textbf{sub-blocks}, as shown at the bottom of Figure~\ref{fig:irregularbuf}. 
The number in each sub-block indicates the target position for the transformation. This sub-block method offers three benefits: (1) it increases the granularity of data transfer, (2) it significantly decreases the cost of the mapping matrix, and (3) it enhances the temporal and spatial locality of the associated data transfer.

Consider a case where the number of processes is $P$, the number of rows and columns in the wavefunction matrix is $r$ and $c$, and $BlockSize = \lfloor r / P \rfloor$. 
In this case, $re \equiv r \mod P$, and there are $(re * c)$ sub-blocks with the size of $(BlockSize + 1)$ and $\left[ (P - re) * c \right]$ sub-blocks with the size of $BlockSize$. 
Figure~\ref{fig:irregularbuf} shows an example with $r=10$, $c = 4$, $P = 4$, and $BlockSize=2$.

Our method starts by dividing the column-partition matrix into two sections as indicated by the dashed line in Figure~\ref{fig:irregularbuf}, ensuring that sub-blocks within each region are of uniform size. 
We then exchange sub-blocks within the same part to create larger sub-blocks, further\textbf{ increasing the granularity of computation}, which is stated in Step \ding{172} of Figure~\ref{fig:irregularbuf}. 
The exchange procedure is straightforward since the target position of each sub-block is reachable to make them sequential in the region.
Following that, we reposition the enlarged sub-blocks to ensure their continuity across the entire region, as demonstrated in Steps \ding{173}, \ding{174}, and \ding{175} of Figure~\ref{fig:irregularbuf}.

In this sub-block transformation scheme, storing the whole mapping matrix is highly redundant, instead, we generate the sub-block mapping index on-the-fly. 
In our experiments, we observe that the computing pattern of this mapping is highly regular, which means we can distribute the mapping workload to each CPE almost evenly.
As a result, we eliminate the mapping matrix without sacrificing any performance.

After MPI\_Alltoallv, each process holds specific rows of the entire matrix. 
We then perform the in-place column rearrangement, which also requires mapping and buffer matrices in the original version. 

\subsubsection{\textbf{Performance and Analysis}}

In the column partition, each process holds a matrix with $r$ rows and $c$ columns. 
Before \texttt{MPI\_Alltoallv}, the mapping matrix and the temporary buffer have the same size. 
Each element in the matrix is represented using double-precision floating-point format, which requires 8 bytes per element. Therefore, the memory requirement for the buffer is $8rc$ bytes.
Regarding the mapping matrix, it can be stored in an integer format, which occupies $4rc$ bytes in memory.
Consequently, the mapping matrices and buffers occupy a total memory of approximately $24rc$ bytes.
For the silicon system with 4096 atoms (Si$_{4096}$), which contains $4096 \times 2 = 8192$ wavefunctions (each silicon atom has four valence electrons, and two valence electrons correspond to a wavefunction), in this case, when using 64 computational grids (CGs) to run the simulation, we have:
\[
c = \frac{8192}{64} = 128 \quad \text{and} \quad r = 592704.
\]
The memory size saved by our aforementioned methods is:
\[
\frac{592704 \times 128 \times 24}{1024 \times 1024} \approx 1736 \text{ MB}.
\]

Meanwhile, our optimization also improves the efficiency of data transformation as the utilization of memory bandwidth is enhanced by transferring data at the sub-block granularity. For example, when 1024 CGs are used to perform data transformation on Si$_{4096}$, we achieve a 14.2x speedup compared with the original version.

\subsection{Communication Optimization}
\label{subsec:allreduce}
\begin{figure}[htp]
    \centering
    \includegraphics[width=\linewidth]{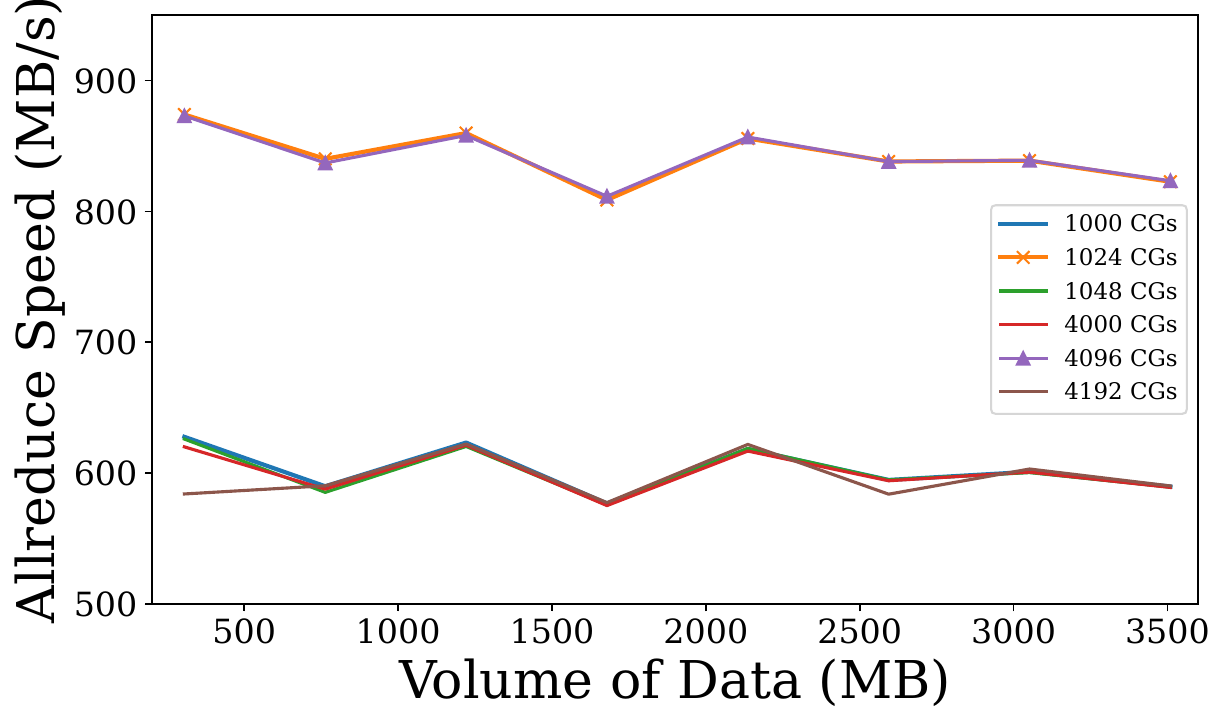}
    \caption{Performance of MPI\_Allreduce regarding different communication volumes on the Sunway supercomputer.}
    \label{fig:fluc}
\end{figure}

It has been documented that MPI collective communication (e.g., \texttt{MPI\_Allreduce}, \texttt{MPI\_Bcast}, etc.) experiences \textbf{performance fluctuations} on some supercomputers~\cite{jia2019parallel}.
In other words, the performance of network bandwidth varies with the volume of data being transferred. 
A similar phenomenon has been observed on the new Sunway supercomputer, as depicted in Figure~\ref{fig:fluc}.
Additionally, a critical observation derived from our experiments suggests that the performance of \texttt{MPI\_Allreduce} on the new Sunway supercomputer is markedly \textbf{superior when the number of participating processes is an exponential power of 2}.

Therefore, we propose a multistage Allreduce to mitigate network performance fluctuations and optimally leverage the communication characteristics of the new Sunway supercomputer.

\subsubsection{\textbf{Notation and Prerequisite}}

We use a predefined parameter $C$ and arrange the processes into a two-dimensional grid, wherein processes in the same row are segmented into a row domain, and likewise, processes in the same column are segmented into a column domain.
For each process, a pair of $i, j$ indicates that the process is situated in the $i-th$ row domain and the $j-th$ column domain. 
The entire matrix is partitioned into $C$ blocks , named $B_{0}, B_{1}, \dots, B_{C-1}$.
These $C$ blocks correspond to the first $C$ processes (having ranks from 0 to C-1) within the row domain.

\subsubsection{\textbf{Algorithm and Implementation}}
\begin{algorithm}[htb]
    \caption{Algorithm of Multistage Allreduce.}
    \label{alg:allreduce}
    \begin{algorithmic}[1]
            \State $rank$: mpi rank in the global communication domain
            \State $P$: number of processes in the global domain
            \State $C$: parameter, each row domain has $C$ or $C+1$ processes
            \State $m \gets P \bmod C$ 
            \State $n \gets rank*(C+1)$
            \If {$rank < n$}
            \State $i\gets \lfloor rank/(C+1) \rfloor $ 
            \State $j\gets rank\bmod (C+1)$ 
            \Else 
            \State $i\gets \lfloor (rank-n)/C \rfloor +m$ 
            \State $j\gets (rank-n)\bmod C$
            \EndIf
            \State Divide the matrix into $C$ blocks, named $B_{0}, B{1} \ldots B{c}$
            \For {$k\in[0,C)$} \Comment{Stage \ding{172}}
            \label{alg:stage1}
            \State {Process with $j = k$ reduce block $B_{k}$ from row domain} 
            \EndFor
            \If {$k\in[0,C)$} \Comment{Stage \ding{173}}
            \label{alg:stage2}
            \State {Execute MPI\_Allreduce in column domain for Block $B_{k}$}
            \EndIf
            \For{processes in the same row} \Comment{Stage \ding{174}}
            \label{alg:stage3}
            \State {Exchange data block with the row domain processes to obtain the entire matrix.} 
            \EndFor
    \end{algorithmic}
\end{algorithm}

In stage \ding{172}, as illustrated in lines 14 to 16 of Algorithm~\ref{alg:allreduce}, processes within the same row domain execute \texttt{MPI\_Reduce} $C$ times to reduce the corresponding block, processing one block per iteration.
Consequently, the process with rank 0 in the row domain obtains the reduced result of block $B_0$, 
while the process with rank 1 obtains the reduced result of block $B_1$, 
continuing in this manner for subsequent processes.
It can be observed that for the processes in the row domain, 
each retains a portion of the reduced result of the matrix (except for extra processes where $j \geq C$).
In other words, $B_0, B_1, \dots, B_{C-1}$ together hold the reduced result of the entire row domain.
As for extra processes, they just send data to other processes with $j < C$ at each \texttt{MPI\_Reduce}, and they don't need to receive any data.
In stage \ding{173} (lines 17 to 19 of Algorithm~\ref{alg:allreduce}), processes within the same column domain $j$ (i.e., processes with identical $j$ values) all retain the same block $B_j$ ($j \geq 0 \quad \text{and} \quad j < C$). 
An \texttt{MPI\_Allreduce} operation is then performed over the column domain, ensuring that every process within column domain $j$ holds the complete reduced result for block $B_j$. 
For extra processes with $j < C$, they are not involved in this stage.
In stage \ding{174} (lines 20 to 22 of Algorithm~\ref{alg:allreduce}), it is evident that the combination of blocks $B_0, B_1, \dots, B_{C-1}$ represents the entire \texttt{MPI\_Allreduce} result. 
Therefore, at this stage, data is exchanged within the row domain via \texttt{MPI\_Alltoallv}.

This approach takes full advantage of the fact the performance of the MPI collective communication is significantly better when the number of participating processes is an exponential power of 2.

We remark that although extra processes (with $j \geq C$) obtain slightly less communication workload, there will be only $P \mod C$ row domains containing an extra process (given that $P \gg C*C$). And our numerical results demonstrate a satisfactory workload balance.

\subsubsection{\textbf{Performance and Analysis}}

\begin{figure}[htbp]
    \centering
    \includegraphics[width=\linewidth]{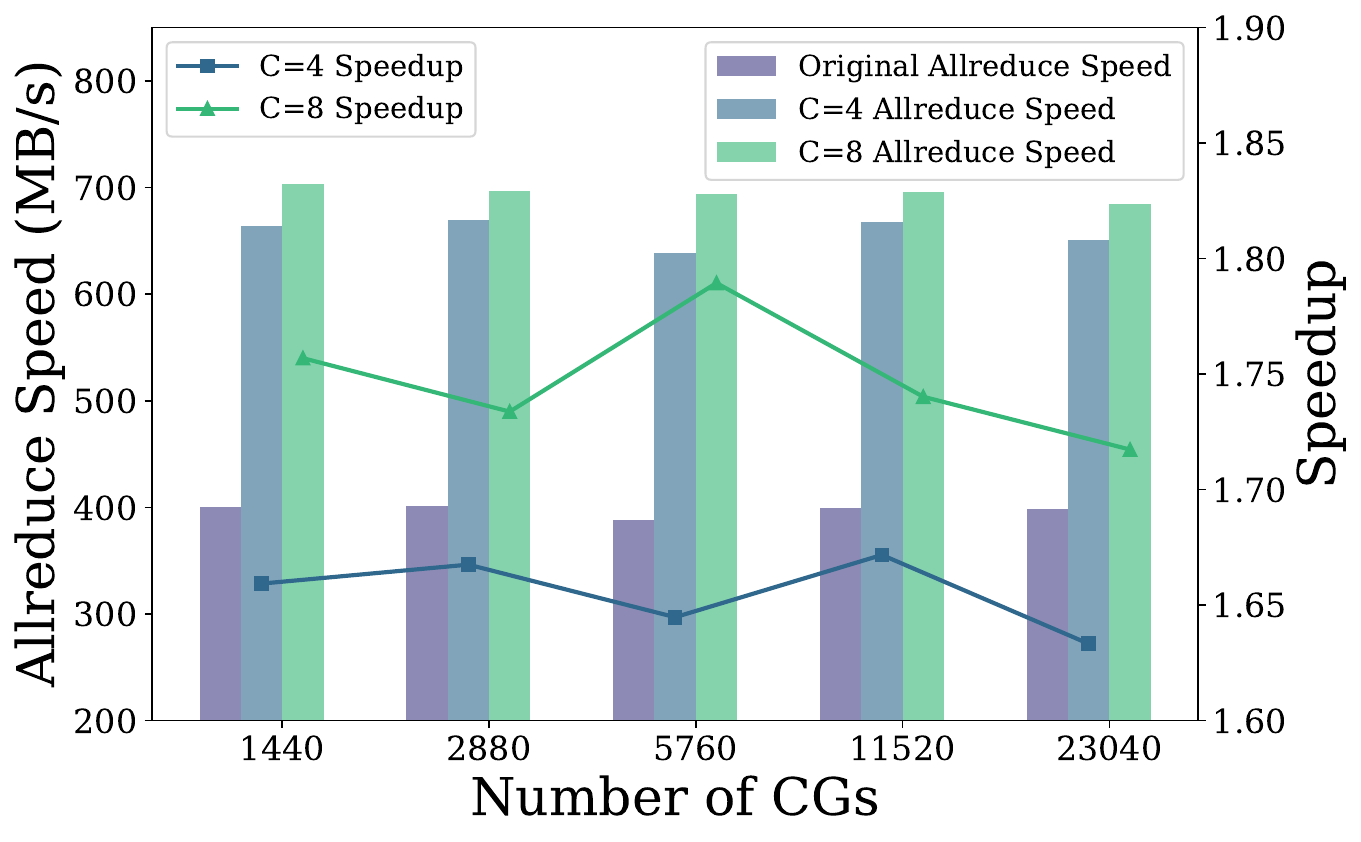}
    \caption{Allreduce performance speedup with the same communication volume and different processes.}
    \label{fig:comp}
\end{figure}

\begin{figure*}[htbp]
    \centering
    \includegraphics[width=\linewidth]{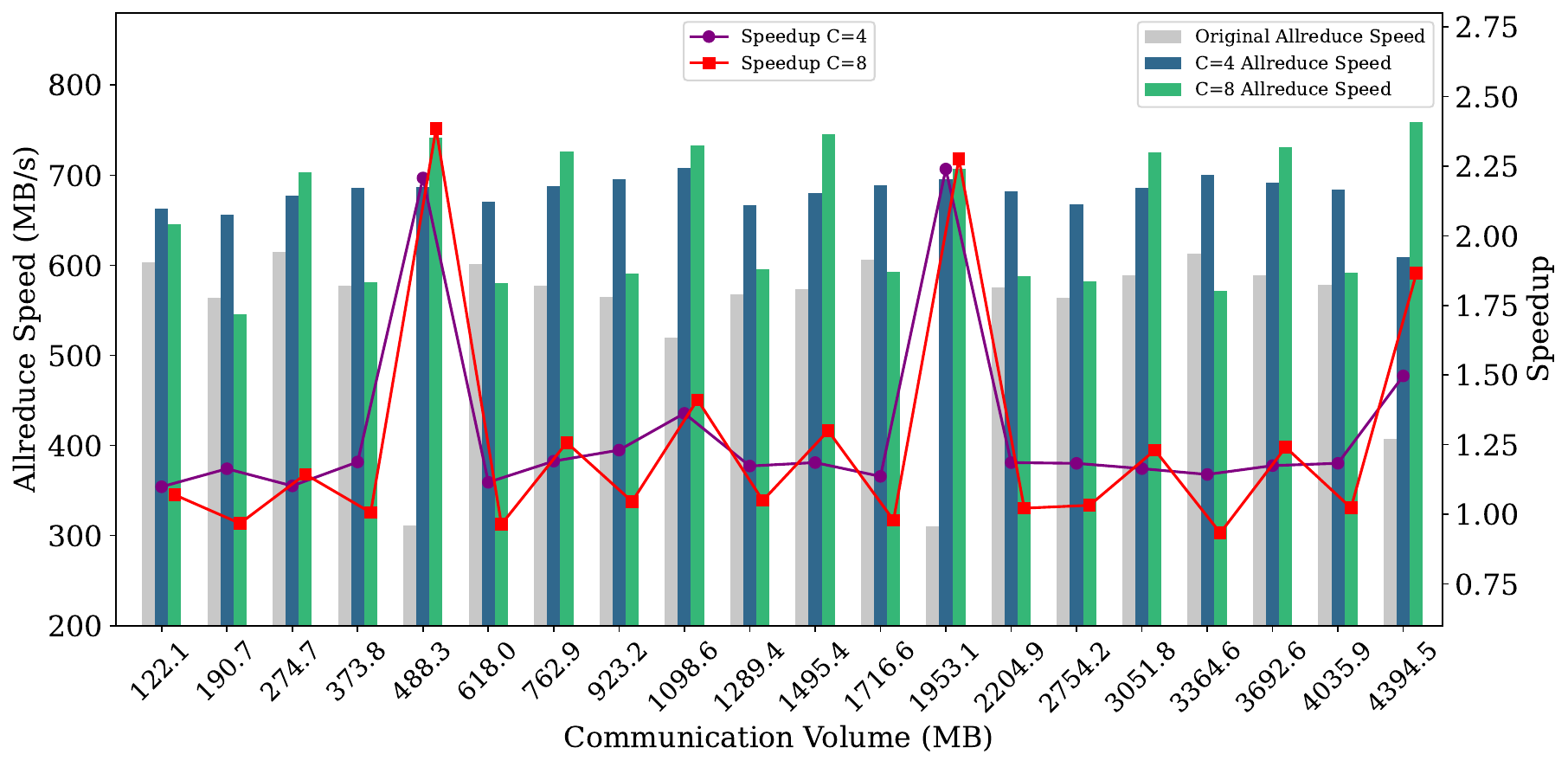}
    \caption{Allreduce performance speedup with different communication volume and the same processes.}
    \label{fig:allreduce_volume}
\end{figure*}

We evaluated our multistage method in the simulation of a graphene system with 11,520 atoms using processes of different sizes. The results in Figure~\ref{fig:comp} show that our implementation is 63.3\% to 67.1\% faster when $C=4$ and 71.7\% to 78.9\% faster when $C=8$. 
To better illustrate the benefits of our optimization, we also evaluated the performance of our algorithm with different communication volumes using the same 4000 processes. 
As shown in Figure~\ref{fig:allreduce_volume}, the stepwise communication reduction algorithm introduced in this manuscript exhibits enhanced performance in the vast majority of instances when $C=4$, achieving an acceleration of up to approximately 2.5 times under optimal conditions.
Although the performance gains at $C=8$ are comparatively modest, they still surpass those of the baseline approach in nearly all cases.

We emphasize that our proposed method is highly adaptable, accommodating any number of processes, and exhibits superior performance when the number of processes is a power of 2.

\subsection{Granularity-Aware Parallelization}
\label{subsec:syevd}

One of the most time-consuming procedures in PWDFT is computing eigenvalues and eigenvectors using Syevd in ScaLAPACK. 
The performance of ScaLAPACK is governed by two factors: (1) the peak performance for computing on each process and (2) the network communication bandwidth. 
It has also been reported that the scalability of ScaLAPACK is limited on modern supercomputers because the workload partitioned in each process is scarce and frequent, and time-consuming communication dominates the performance~\cite{banerjee2018two}. 
We evaluated the Syevd implementation for Si$_{8192}$ using different numbers of processes on the new Sunway supercomputer.
The results presented in Table~\ref{tab:syevd} reveal that the time to solution may even increase as the number of processes grows.
In this case, Syevd performs best when the number of involved processes equals 1024.

% \begin{table}[htp]
% \centering
% \caption{Syevd fo Si$_{8192}$ with different processes}
% \label{tab:syevd}
% \setlength{\tabcolsep}{3mm}{
% \begin{tabular}{@{}cccc@{}}
% \toprule
% ProcNum & RowProcNum & ColProcNum & Syevd Time (s)  \\ \midrule
% 1000 & 8    & 125  & 698.41   \\
% 1024 & 4    & 256  & 210.84   \\
% 2048 & 32   & 64   & 439.23   \\
% 2056 & 257  & 8    & 21349.01 \\
% 4096 & 256  & 16   & 619.06   \\
% 4112 & 16   & 257  & 9858.55  \\ \bottomrule
% \end{tabular}}
% \end{table}

\begin{table}[htp]
\centering
\caption{Syevd time for Si$_{8192}$ with different processes.}
\label{tab:syevd}
\setlength{\tabcolsep}{2mm}{
\begin{tabular}{@{}ccccccc@{}}  \\ \toprule
% \toprule
ProcNum &  1000 & 1024 & 2048 & 2056 & 4096 & 4112 \\ \midrule
Time (s) &  698 & 210 & 439 &  21349 & 619 & 9858 \\ \bottomrule
\end{tabular}}
\end{table}

Therefore, adopting all processes to perform Syevd may suffer performance degradation, and it's sufficient to improve it by simply adopting fewer processes.
To address this problem, we propose a granularity-aware parallelization scheme, which enables PWDFT to run with an optimal process number during different execution steps.
The optimal number of processes is determined by considering various factors, such as computational intensity, communication volume, hardware computational capacity, and network bandwidth available on the new Sunway supercomputer.

As discussed in Section~\ref{subsec:allreduce}, MPI communication benefits from network topology and MPI implementation when the number of processes is a power of 2.
This effect is even more pronounced in the Syevd routine, as we observe significant time-to-solution changes when the number of processes changes from 1000 to 1024, 2048 to 2056, and 4096 to 4112, as shown in Table~\ref{tab:syevd}.
Thus, there is a minimum threshold for the number of processes for a given atomic system, noted as $P_{min}$.
We have developed a C++ library for memory analysis of PWDFT (discussed in Section~\ref{subsec:memory_monitor}), which allows us to analyze memory usage and determine the threshold for the number of processes.
For a silicon atomic system, one atom corresponds to two wavefunctions, the minimum granularity for FFT and matrix multiplication.
Therefore, the maximum number of processes is twice the number of atoms (noted as $A$).

The optimal number of processes (noted as $P_{opt}$) for Syevd falls within the following range: $P_{opt} \in [P_{min}, 2*A] \land P_{opt} = 2^{k}, k \in \mathbb{N}^+$.
Based on this, we can determine the optimal process number from the candidates by balancing computing intensity and communication costs.
For other parts of the PWDFT, such as the calculation of the pseudopotential, the increasing number of processes leads to a shorter time to solution.
Within the range of $[P_{min}, 2*A]$, more processes indicate faster execution.

\subsection{SWUC-assisted Heterogeneous Multi-core Programming}
\label{subsec:cpe}
The extensive floating-point performance on the new Sunway supercomputer is mainly provided by CPEs.
Each CPE has a very limited (256 KB) scratch-pad memory (SPM) that can be configured as local data memory (LDM) or Local Data Cache (LD Cache).
However, effectively harnessing the CPEs to maximize the acceleration of PWDFT proves to be a challenging task.
It requires writing extensive code via the Athread library and working within the confines of the limited LDM.

Fortunately, a new programming framework called ShenWei Universal C/C++ (SWUC)~\cite{cao2022design} has been proposed to simplify programming with CPEs.
Thanks to several new attributes and compiler directives, writing code running on CPEs and MPE can be simply achieved by calling lambda expressions, in which we divide data for each CPE and specify the code to be executed.
We adopt the framework in the calculation of pseudopotential, wavefunctions, charge density, and energy.
We divide the data according to the layout of CPEs, batch data, and transfer it to the LDM of CPEs, and obtain the final results via a lambda function.

However, for some more complex computing scenarios, SWUC is not yet applicable.
In such cases, we have to use the Athread library, the most primitive interface to use CPEs, to accelerate PWDFT.
The numerical result of the SWUPC-assisted CPE programming will be discussed in Section~\ref{subsec:cpe_speedup}.

\subsection{Other Optimization and Tool}
\label{subsec:tools}
In this subsection, we briefly describe our minor contributions to PWDFT on the new Sunway supercomputer.

\begin{figure}[htbp]
    \centering
    \includegraphics[width=\linewidth]{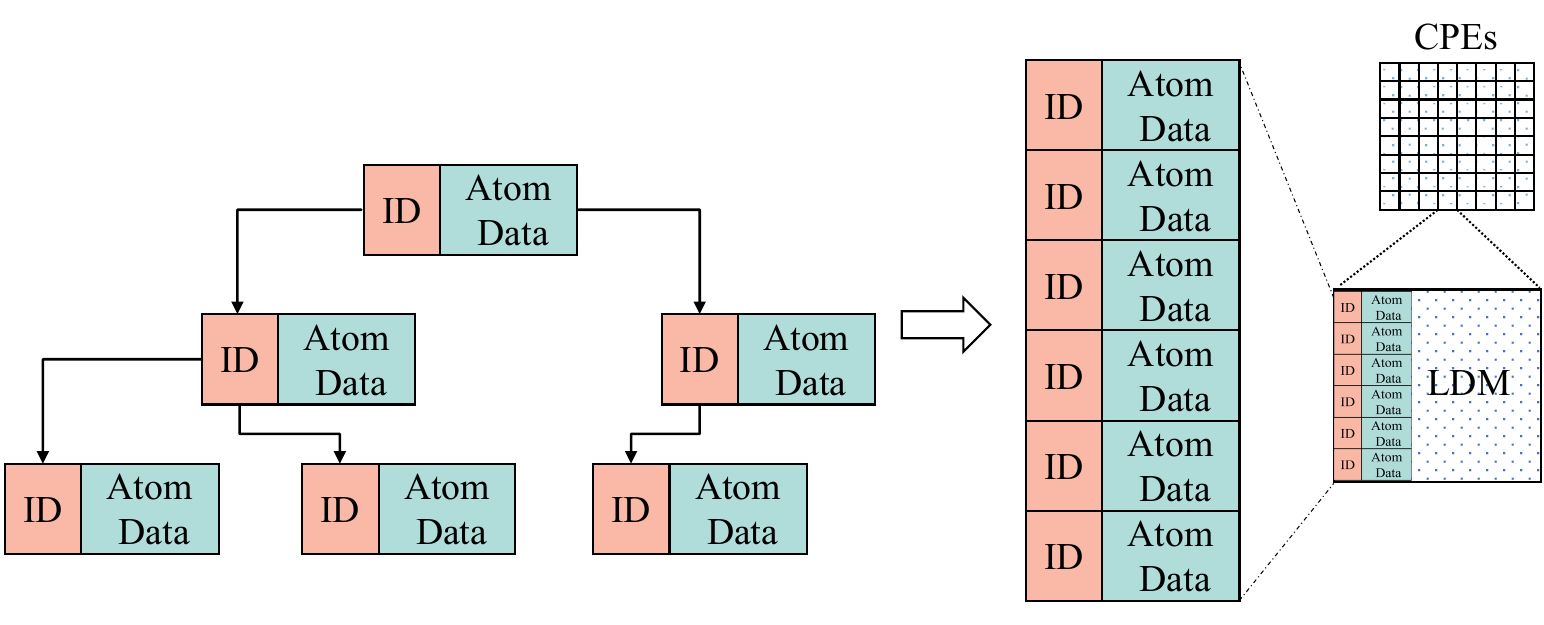}
    \caption{Data Structure Optimization of Pseudopotential.}
    \label{fig:Map2AoS}
\end{figure}

\textbf{Data Structure Optimization:}
When constructing the PW Hamiltonian, it's necessary to look up the pseudopotential for a given atom. However, the original implementation of PWDFT, which targets general-purpose CPUs, used std::map from the C++ Standard Template Library to perform the look-up. 
This method suffers from poor performance due to its indirect access, especially when the system size increases. 
Additionally, the CPEs on SW26010Pro are unable to allocate from the main memory shared across cores, making std::map inaccessible on CPEs and hindering pseudopotential lookup acceleration.
Considering that there are typically no more than 200 different types of atoms, to address these issues, we replace std::map with a small array. 
By indexing the array with the atom number, we are able to acquire the pseudopotential with one direct access, resulting in significant performance improvements, as shown in Figure~\ref{fig:Map2AoS}. 
This new mechanism is also more CPE-friendly, as the LD Cache makes loading the pseudopotential efficient in the case of direct indexing. Compared to the previous design, this method boosts the performance of the PW Hamiltonian step by a factor of 10.

\textbf{Runtime Memory Usage Monitor:}
\label{subsec:memory_monitor}
To obtain a more detailed understanding of memory usage, we implemented a runtime memory monitor library for PWDFT.
Memory usage can be categorized into three main types: global memory space for global variables, temporary memory space allocated for temporary variables in a function call, and memory space to store DFT information.
We ignore the first type, as it is relatively small in PWDFT.
We do not need to pay special attention to the second type of memory usage if no memory bound occurs in the function.
We consider all variables for DFT information and calculate the used memory size.
For memory-bound functions, we enumerate the large temporary variables that occupy significant memory space and determine the total size of the occupied space, including DFT information.
The memory usage information is outputted whenever there is a substantial change in memory usage.

\subsection{Portability}
Our optimizations can significantly accelerate PWDFT and extend the limits for DFT calculation.
The techniques in this work can be readily applied to other platforms, except for the CPE optimizations.
Additionally, these optimizations are relevant and serve as a reference for DFT software across various platforms.
Specifically, the refactoring of the pseudopotential algorithm and memory access optimization can be easily ported to other DFT software on platforms such as GPU or many-core X86/Arm CPU supercomputers without significant code changes.
Furthermore, communication optimization and the granularity-aware parallelization scheme can be applied to other software on the new Sunway supercomputer and other computers with similar characteristics~\cite{jia2019parallel,banerjee2018two}.
Our optimization with CPE in Section~\ref{subsec:cpe} is specific to the Sunway platform and is highly beneficial for other research conducted on the Sunway supercomputer.

\section{Numerical Results and Analysis}

\subsection{Setup of the Test Physical Systems and Testing Environment}
\begin{figure}[htp]
    \centering
    \includegraphics[width=\linewidth]{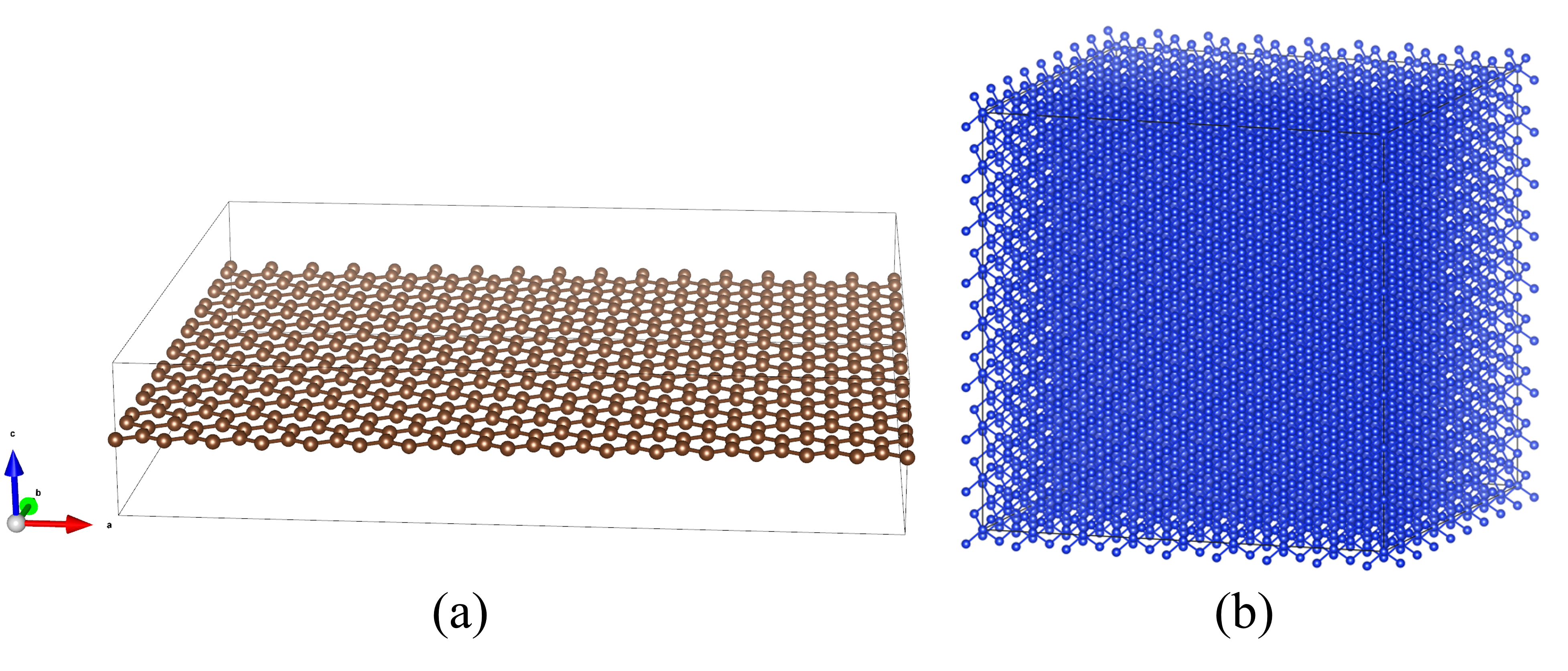}
    \caption{Testing systems: (a) graphene system and (b) bulk silicon system.}
    \label{fig:test_system}
\end{figure}
The testing systems include two parts: graphene (labeled as G) and cubic silicon systems (labeled as Si), as shown in Figure~\ref{fig:test_system}.
These systems are denoted respectively as G$_{11520}$, G$_{16384}$, Si$_{64}$, Si$_{216}$, Si$_{512}$, Si$_{1000}$, Si$_{4096}$, and Si$_{8192}$.
The corner marks indicate the number of atoms in each system.
Notably, our parallelization approach is well-suited for handling large-scale quantum simulations, making it applicable to other complex physical systems such as amorphous silicon~\cite{pedersen2017optimal,xu2023velocity}.

% ---- Delete for the Euro-Par submission ----
% As mentioned above, computing intensity will be increased when the value of $E_{cut}$ increases. 
In our experiment, the kinetic energy cutoff $E_{cut}$ is set to 5 Hartree unless otherwise specified.
This value has been found sufficient for achieving convergence in silicon crystal calculations~\cite{cances2024modified}. 
We use the same $E_{cut}$ in graphene systems for consistency.
Our optimization approach has good effects for different values of $E_{cut}$.
All numerical tests are performed on the new Sunway supercomputer, and each MPI process is applied to one CG (including 1 MPE and 64 CPEs) with separated 16 GB of memory. We consider the time per outer SCF iteration as the time to solution.

% ---- Delete for the Euro-Par submission ----
% Meanwhile, some general scientific computing libraries such as ScaLPACK and LAPACK have been adapted to the new Sunway supercomputer, which is used for the calculation of GEMM, FFT, and so on.

\subsection{Strong Scalability}
To evaluate the strong scalability of Si$_{4096}$ and Si$_{8192}$, we use the parallel efficiency defined as
\[
\text{efficiency} = \frac{\text{Speedup}}{\text{Multiple of CGs}}.
\]
The speedup is compared with the minimum number of CGs (64 CGs for Si$_{4096}$ and 512 CGs for Si$_{8192}$) capable of performing the calculation.
We use 64 to 2048 CGs when simulating Si$_{4096}$ to test the strong scalability of PWDFT and 512 to 4096 CGs for Si$_{8192}$.

\begin{figure}[htp]
    \centering
    \includegraphics[width=\linewidth]{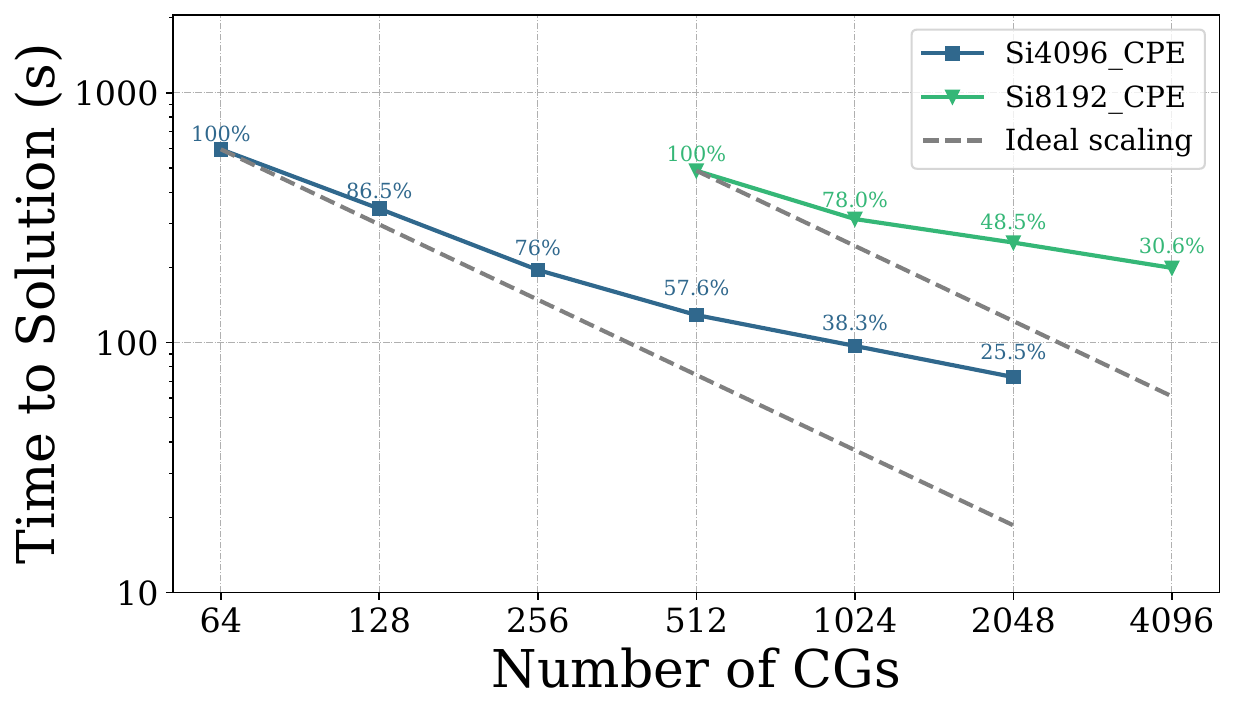}
    \caption{Strong scaling of PWDFT.}
    \label{fig:strong}
\end{figure}

The results are shown in Figure~\ref{fig:strong}.
From the figure, we can see that, when scaling up to 2048 CGs (Si$_{8192}$), PWDFT maintains a parallel efficiency of 48.5\%, which is quite acceptable for PW-based DFT calculations.
However, the parallel efficiency is not satisfactory when scaling up to half the number of atoms.
Even so, we must note that such mediocre parallel efficiency is already much better than the original version and other PW-based DFT software.
In Section~\ref{subsec:syevd}, due to the poor scalability of Syevd, the optimal process number is always selected in our granularity-aware parallelism for Syevd in the range of $P_{opt} \in [P_{min}, 2*A] \land P_{opt} = 2^{k}, k \in \mathbb{N}^+$.
Hence, when the number of processes reaches a sufficient threshold to achieve the best performance for Syevd, it remains unchanged.
Considering that Syevd often occupies a significant portion of the total computation time, it is reasonable to conclude that PWDFT does not exhibit excellent strong scalability and parallel efficiency.
Additionally, massive global communication, as discussed in Section~\ref{sec:massive_comm}, also partially contributes to this inefficiency.

\begin{figure}[htp]
    \centering
    \includegraphics[width=\linewidth]{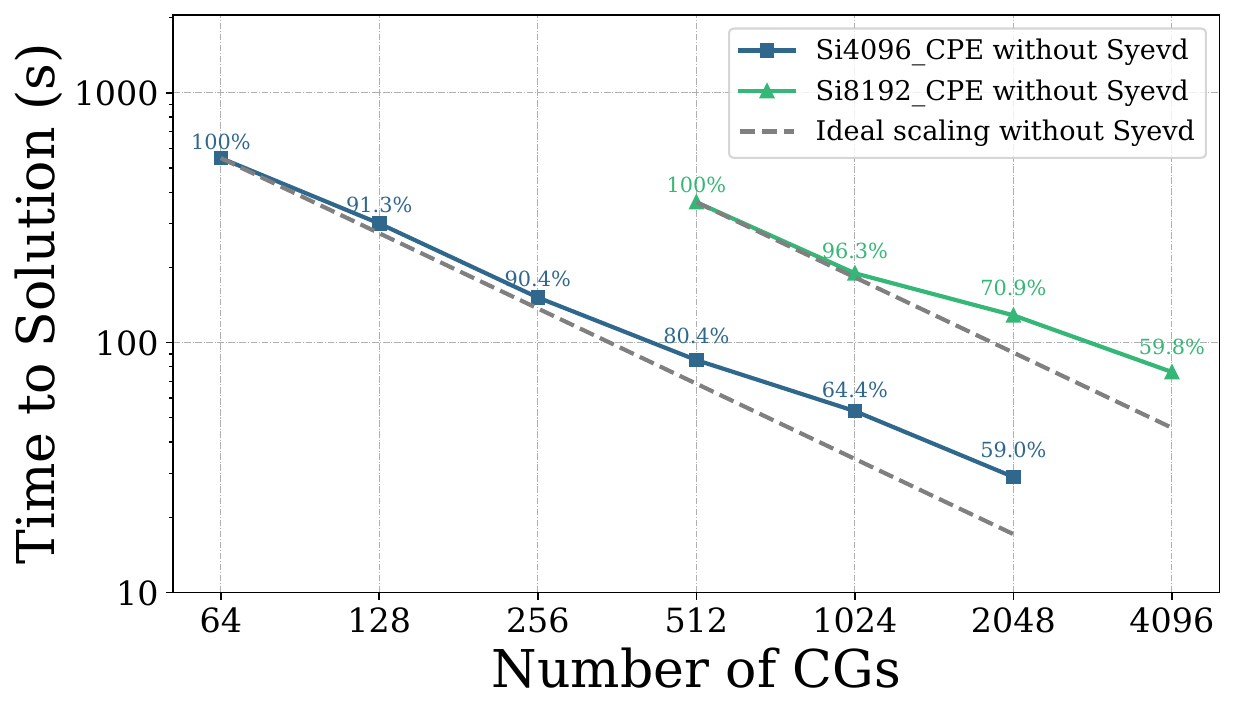}
    \caption{Strong scaling of PWDFT without syevd.}
    \label{fig:strong_with_out_syevd}
\end{figure}

As is shown in Figure~\ref{fig:strong_with_out_syevd}, the result of the above systems excluding Syevd time shows a much better and reasonable parallel efficiency as over 59\% of the parallel efficiency is achieved when the number of CGs reaches half of the atom's number.
Moreover, under such circumstances, massive global communication starts to play a dominant role when the number of CGs is greater than half of the atom's number, which prevents the code from scaling further.

\subsection{Weak Scalability}
Our method offers a significant reduction in memory cost during the calculation steps of PWDFT. 
As a result, we can efficiently study a much larger physical system using far fewer computing resources. 
In our testing of weak scalability for Si$_{64}$, Si$_{216}$, Si$_{512}$, Si$_{1000}$, Si$_{4096}$ and G$_{16384}$, we utilize one-quarter of the atom number as the process number. 
This approach enables us to optimize computational efficiency while maintaining high accuracy in our calculations.

\begin{figure}[htbp]
    \centering
    \includegraphics[width=\linewidth]{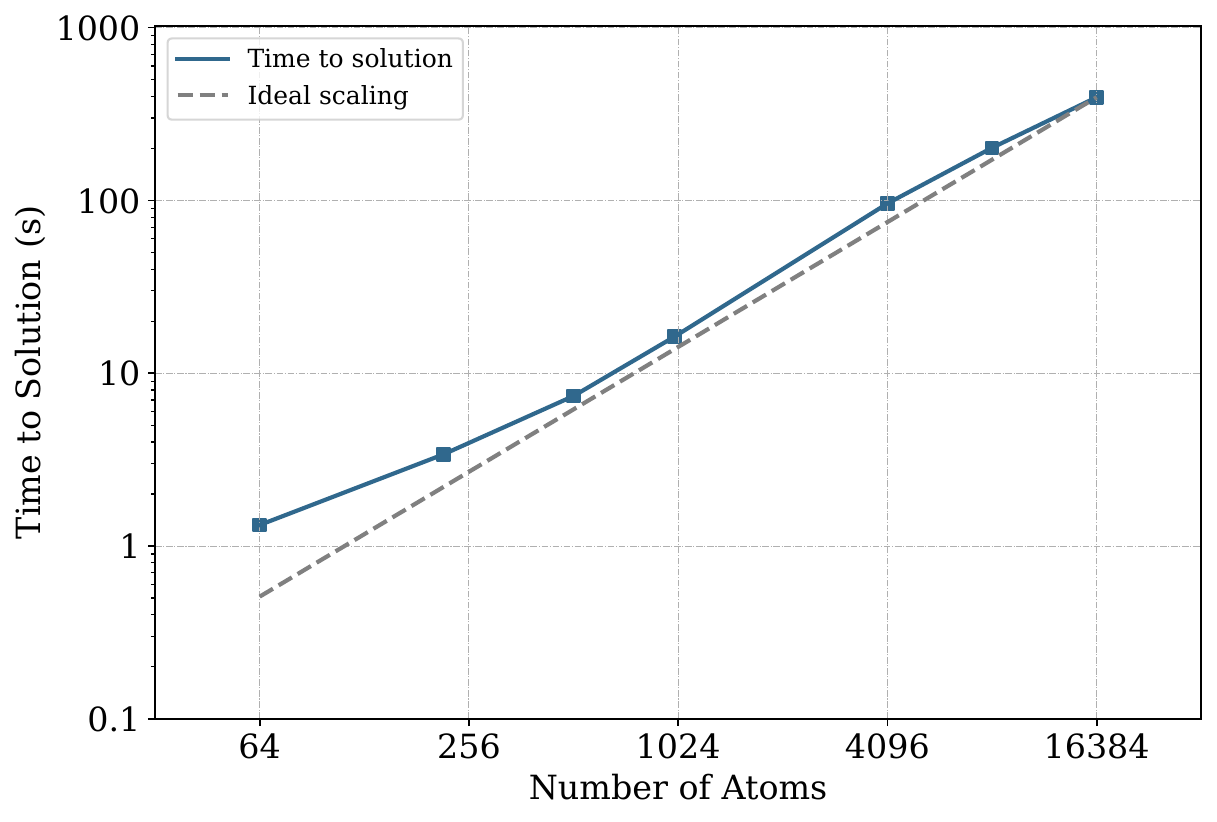}
    \caption{Weak Scaling of PWDFT w.r.t. the number of electrons.}
    \label{fig:weak}
\end{figure}

The weak scaling of PWDFT for simulating systems consisting of 64 to 16,384 atoms is shown in Figure~\ref{fig:weak}, from which we can see that PWDFT exhibits excellent weak scaling properties.
Since the computational complexity of PWDFT is $O(N^3)$, and the number of CGs is set to a quarter of atoms number, the ideal scaling should be $O(N^2)$, which is represented by a dotted line in Figure~\ref{fig:weak}.

As shown in Figure~\ref{fig:weak}, in the case of small systems, computing resources are not the dominant factor, and the cost of communication becomes relatively more significant, making scalability a challenge. 
Therefore, it is difficult to ignore the impact of communication costs, which can limit the ability of the system to scale effectively.
When the atom's number is greater than 1000, the actual execution time is very close to the ideal scaling, which suits our computational complexity well.
Here for G$_{16384}$, the simulation with 4,096 CGs is only 395 seconds, which makes DFT calculation for large systems more practical.

\subsection{Speedup Compared with Master Version}
\label{subsec:cpe_speedup}

\begin{figure}[htp]
    \centering
    \includegraphics[width=\linewidth]{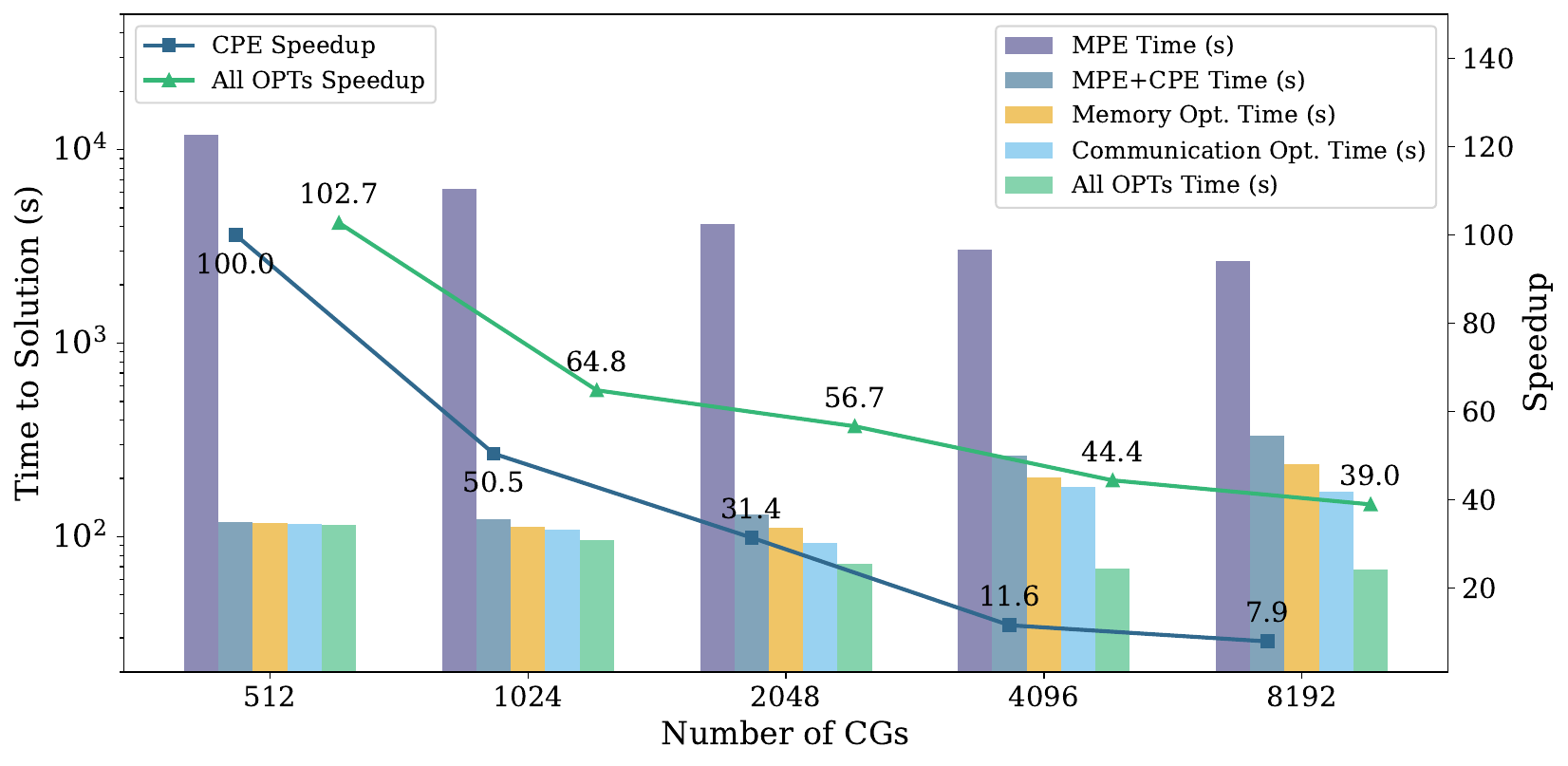}
    \caption{Speedup for Si$_{4096}$ with master version.}
    \label{fig:speedups}
\end{figure}

In Figure~\ref{fig:speedups}, we compare the acceleration effects of using CPE alone and employing all optimization techniques in the simulation of Si$_{4096}$. 
We use the master version's results, which are directly transferred to the Sunway system and rely solely on MPE, as the baseline for our comparison.

The results show that when only adopting CPE to accelerate, the speedup varies from 100.0 to 7.9, and decreases dramatically with the increase of the number of CGs.
For comparison, when utilizing all the mechanisms proposed in Section~\ref{sec:innovations}, the acceleration effect is noticeably enhanced.
Under more common circumstances, the fully optimized version achieves 64.8 times faster than the master version when we adopt one-quarter of the atom number, i.e., 1,024 CGs.

For both versions, the acceleration decreases with an increase in CGs.
This is because an increase in CGs indicates less computational workload, and the cost of communication also rises.

\begin{table}[htbp]
\centering
\caption{Minimum CGs required to calculate the system.}
\label{tab:minCGs}
\begin{tabular}{@{}ccccc@{}}
\toprule
System          & Master Version & Optimized Version  \\ \midrule
Si$_{1000}$     & 512      & 25   \\
Si$_{4096}$           & 512     & 64   \\
Si$_{8192}$           & Out of Memory    & 512  \\
G$_{11520}$           & Out of Memory    & 720 \\
G$_{16384}$           & Out of Memory    & 2048 \\
\bottomrule
\end{tabular}
\end{table}

Meanwhile, memory optimization in Section~\ref{subsec:pseudo} and Section~\ref {subsec:alltoall} significantly extends the limit of the studied system for PWDFT. As shown in Table~\ref {tab:minCGs}, 
The master version requires at least 512 CGs to calculate a silicon system containing 4,096 atoms (Si$_{4096}$). 
However, after our optimization, only 64 CGs are needed. Furthermore, only 512 CGs are required for a silicon system with 8,192 atoms (Si$_{8192}$), which is quite impressive for PW-based DFT calculations.
Consequently, the limit of PWDFT is extended from a physical system of 4,096 atoms to one encompassing 16,384 atoms.

\subsection{Speedup with Different Energy Cutoff}
\label{subsec:ecut}

\begin{figure}[htbp]
    \centering
    \includegraphics[width=\linewidth]{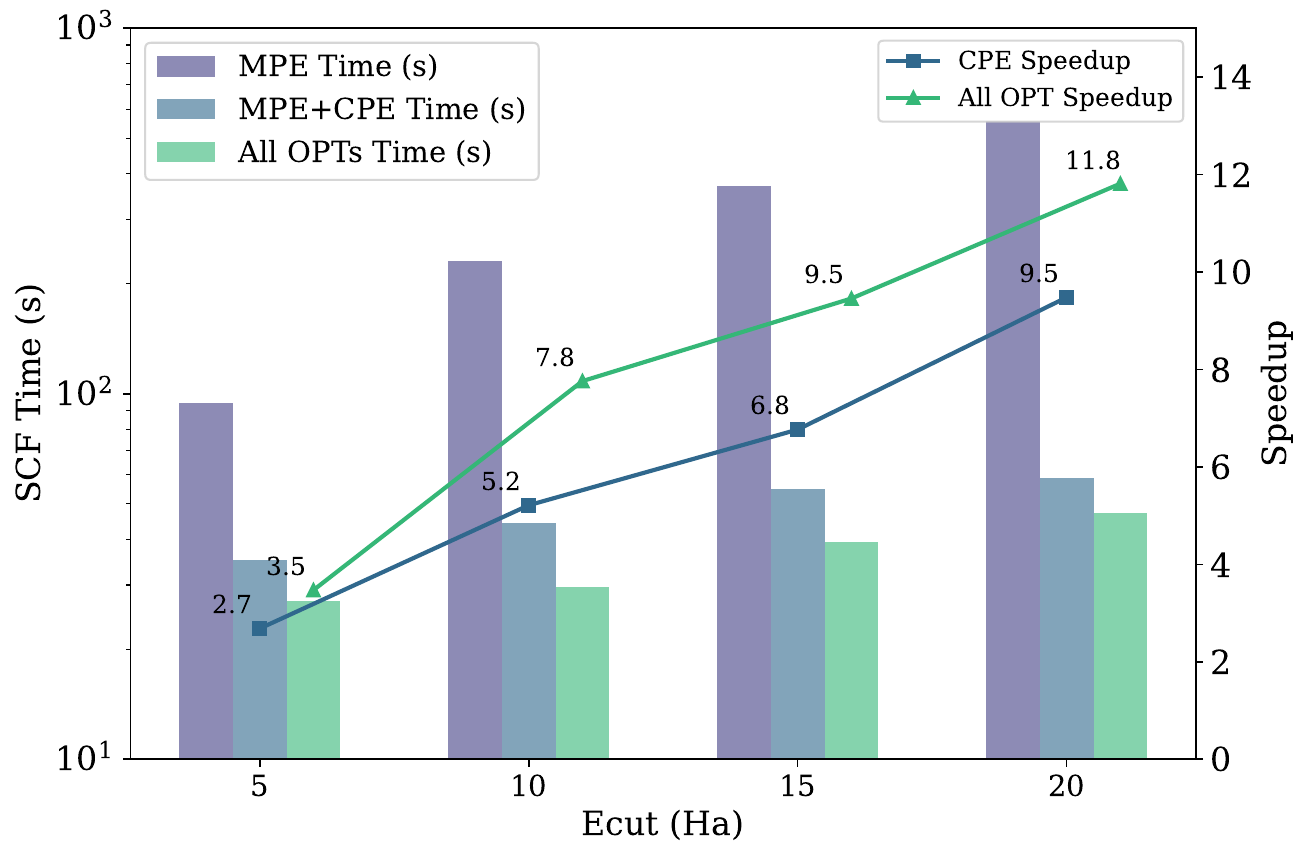}
    \caption{SCF iteration time and speedup comparisons at different $E_{cut}$ values.}
    \label{fig:ecut_impact}
\end{figure}

To comprehensively examine the performance of our optimization strategies across various computational conditions, we conducted additional tests by varying the plane-wave energy cutoff ($E_{cut}$). 
We selected four different $E_{cut}$ values (5, 10, 15, and 20 Ha), representing both our default setting (5 Ha) and more realistic values typically encountered in practical DFT calculations. 
We compared the SCF iteration times of three versions of our implementation: the Master version (MPE, utilizing only the main processing element), the version with additional computing processing elements (MPE+CPE, including slave cores), and the fully optimized version (All OPT), incorporating all optimization techniques proposed in this work. 
The results are summarized in Figure~\ref{fig:ecut_impact}.

From Figure~\ref{fig:ecut_impact}, we observe the following:

\begin{itemize}
    \item At all tested energy cutoffs, both the MPE+CPE and All OPT versions significantly outperform the Master version, clearly demonstrating that utilizing slave cores effectively accelerates SCF calculations.

    \item The speedup of the MPE+CPE version relative to the Master version improves as the energy cutoff increases, achieving approximately 2.7$\times$, 5.2$\times$, 6.8$\times$, and 9.5$\times$ at 5, 10, 15, and 20 Ha, respectively. 
    This trend underscores the effectiveness of parallelization using additional computing elements as computational intensity increases.

    \item More importantly, the fully optimized All OPT version, which includes all optimizations discussed in this paper, consistently delivers superior performance.
    Specifically, it achieves a speedup of approximately 3.5$\times$ at 5 Ha, which steadily rises to approximately 7.8$\times$, 9.5$\times$, and 11.8$\times$ at 10, 15, and 20 Ha, respectively. 
    This substantial improvement further validates the robustness and practical applicability of our proposed optimization strategies.

\end{itemize}

These results confirm that the optimization methods proposed in this study are effective across a broad range of energy cutoffs, ensuring that our conclusions remain valid and relevant under realistic computational scenarios commonly encountered in large-scale PW-based DFT simulations.

\section{Discussion}
% 分析一下目前程序瓶颈，热点等问题，并且给出当前情况下的限制（来自内存？来自通信？），并增加计算访存方面的分析，比如现在优化后的PWDFT是memory-bound or compute-bound，在之后的工作中如何针对性地在硬件或软件层面提升他的效果。可以从计算访存比的角度讨论优化前/优化后的PWDFT对于神威平台的适应性，并声明经过精心的优化，PWDFT变得更加适应新神威这种超高计算访存比的架构。

One of the hot spots in PWDFT is solving the eigenproblem distributively by calling Syevd. 
While other dominant functions have been optimized and parallelized, the Syevd function does not benefit significantly from an increase in the number of processes due to communication constraints. 
This has been a consistent finding in our research.

In terms of computational ability, PWDFT is still memory-bound for the simulation of larger systems on the new Sunway supercomputer. 
Memory bottleneck occurs in the storage of pseudopotential and Hamiltonian matrices. 
To our knowledge, our work has already reached the boundary of PW-based DFT on the new Sunway supercomputer.
To further expand the capabilities of PW-based DFT software on the new Sunway supercomputer, one must either (1) create novel algorithms with reduced memory complexity or (2) develop numerical libraries that align with the large memory scheme of the new Sunway supercomputer. 
It is important to emphasize that this work does not primarily focus on these two aspects.

Our memory optimization approach reduces random memory access, thereby mitigating the negative impact of data movement and making PWDFT more adaptable to new Sunway’s ultra-high computing-to-memory ratio architecture.

\subsection{Potential Applications of PWDFT-SW}
\label{subsec:future_work}

The optimization techniques presented in this work are primarily developed and evaluated for the Sunway supercomputer; nevertheless, many of these strategies are not inherently limited to the Sunway architecture and can be generalized to other state-of-the-art supercomputing platforms, including GPUs~\cite{v100,a100}, Fugaku~\cite{sato2021co}, and Summit~\cite{wells2016announcing}. 

For instance, the memory accessing optimizations and pseudopotential refactoring schemes, designed to reduce redundant communication and enhance data locality, are directly applicable to GPU-based platforms such as Summit, which benefit significantly from efficient memory management and reduced communication overhead. 
Similarly, the granularity-aware parallelization method can be adapted for other many-core CPU architectures, such as Fugaku, where appropriate resource allocation and optimized communication patterns can greatly enhance computational performance.

Future work will involve porting and tailoring these optimization strategies explicitly for GPU-based and ARM-based supercomputers. 
Such efforts would enable detailed benchmarking across different architectures, highlighting the universality and adaptability of our proposed optimizations. 
Additionally, further investigation into platform-specific tuning and potential new bottlenecks on these architectures would provide valuable insights, thereby broadening the applicability and impact of our methods within the high-performance computing community.

\begin{table*}[htb]
\centering
\caption{List of Different PW-based DFT Software.}
\label{tab:software}
\begin{tabular}{@{}cccccc@{}} \\ \toprule
Software                       & Year      & System                     & Atom  & Machine                                & Scale (cores)       \\ \midrule
Qbox\cite{osti_1469055}    & 2017 &  Gold  & 9K  & Trinity & 300K  \\
Qbox\cite{gygi2006large}       & 2006      & Mo                         & 1K    & BlueGene/L                             & 128K         \\
VASP\cite{2018Large}           & 2018      & Amorphous silicon          & 1K    & Ohio Supercomputer and  Tesla K40 GPU  & not mentioned    \\
PEtot\cite{jia2013analysis}    & 2013      & CdSe                       & 1K    & Cray XK6, IPE                          & 256 GPUs            \\
QE\cite{giannozzi2017advanced} & 2017      & Clycine, H2O               & 768   & Cray XC40                              & 65K          \\
PWDFT\cite{hu2017adaptively}   & 2017      & Si                         & 4K    & Edision, Cori                          & 8K             \\
PWDFT                          & This work & Graphene, Si               & 16K &  New Sunway                         & 32K               \\ \bottomrule
\end{tabular}
\end{table*}

\subsection{Integration with Machine Learning Methods}
Machine learning (ML) techniques have demonstrated significant potential in accelerating and enhancing DFT calculations.
Recent prominent examples include DeepH~\cite{li2022deep,wang2024deeph}, which employs graph-based neural networks to predict DFT Hamiltonians directly from atomic structures, thereby circumventing expensive self-consistent iterations. 
Another noteworthy development is DeepKS~\cite{chen2020deepks,chen2023deepks}, which uses neural-network corrections to improve the accuracy of conventional DFT exchange-correlation functionals by learning from high-accuracy reference data.
Additionally, DeepMD~\cite{wang2018deepmd,zeng2023deepmd} has revolutionized ab initio molecular dynamics simulations by using ML-driven interatomic potentials, significantly enhancing both computational efficiency and scalability. 
The remarkable success of DeepMD, highlighted by the ACM Gordon Bell Prize~\cite{lu202186}, underscores the transformative impact of ML in enabling large-scale molecular dynamics simulations with DFT-level accuracy.
Furthermore, recent generative models such as MatterGen~\cite{zeni2025generative} have also demonstrated promising capability in efficiently generating stable inorganic materials, potentially complementing ML-accelerated DFT calculations.
Additionally, the recent N2AMD framework~\cite{zhang2025advancing}, employing E(3)-equivariant neural networks to predict DFT Hamiltonians for nonadiabatic molecular dynamics, offers a promising direction for further accelerating PWDFT-based simulations of large-scale materials.

Beyond these specific frameworks, general ML strategies like Behler-Parrinello neural networks~\cite{groenenboom2020combined} and Gaussian Approximation Potentials (GAP)~\cite{kocabacs2023gaussian} have also been integrated successfully with DFT to create computationally efficient surrogate models, reducing computational demands while preserving high fidelity.
Integrating such ML-based methods with the optimization strategies proposed in this study holds substantial promise. 
Future investigations could involve combining our parallel optimization methods with ML-based approximations to further enhance efficiency, particularly in large-scale materials simulations. 
Such integration could expand the application of PWDFT to even larger and more complex systems, offering avenues for significant future research.

\section{Related Works}
\subsection{Other Parallel PW-based Software}

There are several software packages based on plane wave available for performing DFT, such as Vienna Ab initio Simulation Package (VASP), Qbox, Quantum ESPRESSO (QE), PEtot~\cite{jia2013analysis}, and PWDFT~\cite{hu2017adaptively}. 
VASP, Qbox, and QE are traditional software mainly used for simulations of medium-size atoms (hundreds of atoms), and PWDFT and PEtot are designed for large-size atoms (thousands of atoms). 
We list the related works in Table~\ref{tab:software}.
It is important to note that the list provided in Table~\ref{tab:software} involves results obtained from different physical systems and computational resources, thus serving primarily as a general reference rather than a rigorous benchmarking.
The current work emphasizes performance optimization tailored specifically for the Sunway architecture, highlighting scalability and capability for handling large-scale systems. 
A detailed comparison under consistent accuracy settings and computational conditions has been reported separately in PyPWDFT~\cite{gao2025pypwdft}.

To optimize plane-wave-based DFT calculation, one way is to port to the latest heterogeneous or homogeneous computing platforms~\cite{jia2013analysis, jia2019parallel, zhao2018porting}, another way is to accelerate at the numerical algorithm level~\cite{hu2017adaptively, jia2019parallel, jiang2022accelerating}.

Recently, thanks to heterogeneous computing systems such as graphics processing units (GPU) and many-core microarchitectures, those traditional software packages have come to rejuvenate after being ported onto the new generation of supercomputers.
In 2006, the winner of the Gordon Bell Prize extended the limit of Qbox on the BlueGene/L supercomputer to 1K atoms, scaling up to 128k cores on 65,536 nodes~\cite{gygi2006large}. In 2017, researchers successfully ran Qbox~\cite{osti_1469055} on 9418 nodes on Trinity with 8,800 atoms. In 2013, Jia~\cite{jia2013analysis} implemented  PEtot on a multi-node GPU machine with x13-x22 speedups over the CPU calculations for a typical 512-atom system. 
% They claimed that their implementation was able to calculate up to 4000 atom systems with enough CPU/GPU units, but they did not really do it. 
Researchers ported VASP to Sunway TaihuLight~\cite{zhao2018porting} and achieved 2.90x to 4.48x speedup, but they only focused on small cases containing 32 Ge atoms and one Ru atom. It is significantly smaller than other works. 

% \todo{error analysis compared to other DFT software}

\subsection{Other Research Works on the New Sunway Supercomputer}
There are a lot of successful works that effectively utilize the massive computing resource of the new Sunway supercomputer~\cite{hu20222, zhao2022ai, shang2022large, wang2022scaling}, for instance: (1) By implementing a novel sparse direct solver PEXSI, DGDFT extended the limit of adaptive local basis DFT calculation to 2.5 million atoms using 35.9 million cores, achieving ~5\% theoretical peak performance~\cite{hu20222}. (2) A brand new computational framework to solve quantum many-body problems by Artificial Intelligence (AI) was proposed and adapted to the Sunway, scaling up to 40 million heterogeneous cores with high efficiency~\cite{zhao2022ai}. (3) Based on the abstraction of application and hierarchical hardware, researchers presented a unified programming model, UniQ, for multiple simulation methods. UniQ can generate execution with high performance without any human involvement. It scales to 399,360 nodes and achieves up to 28.59x speedup over the state-of-the-art framework~\cite{zhang2022uniq}.

\section{Conclusion}
In summary, we have significantly optimized the ability and performance of PWDFT on the new Sunway supercomputer. 
Firstly, we reduce memory usage and improve memory bandwidth utilization by refactoring pseudopotential calculations and in-place data transforming, so that a single CG with 16 GB memory is able to hold much more wavefunctions and a larger Hamiltonian matrix for larger systems. 
Secondly, we accelerate PWDFT by adopting CPEs and granularity-aware parallelization to take full advantage of physical floating-point performance. 
Meanwhile, we also present a multistage Allreduce to optimize network bandwidth utilization. 
As a result, PWDFT can now perform calculations of 4096 atoms using 64 CGs, 720 CGs for 11,520 atoms, and 2,048 CGs for 16,384 atoms.
These optimizations lead to a 64.8x speedup when simulating a silicon system comprising 4,096 atoms and utilizing 1,024 CGs.
Additionally, our experiments demonstrate that our parallel design can achieve good strong scalability, even when utilizing a large number of cores.

\section*{Acknowledgment}
We thank the reviewers of ICPP 2023, IPDPS 2024, Euro-Par 2024, and TPDS for their feedback to strengthen this work.
This work is supported by the Strategic Priority Research Program of Chinese Academy of Sciences (XDB0500102, XDB0450101, XDB1170000), the Laoshan Laboratory (LSKJ202300305), the Innovation Program for Quantum Science and Technology (2021ZD0303306), the National Natural Science Foundation of China (22173093, 22373096, 21688102), the Anhui Provincial Key Research and Development Program (2022a05020052), the National Key Research and Development Program of China (2021YFB0300600), the CAS Project for Young Scientists in Basic Research (YSBR-005), and the Fundamental Research Funds for the Central Universities (WK2320000061).

\bibliographystyle{splncs04}
\bibliography{IEEEabrv,sample-base}

\begin{thebibliography}{10}
\providecommand{\url}[1]{\texttt{#1}}
\providecommand{\urlprefix}{URL }
\providecommand{\doi}[1]{https://doi.org/#1}

\bibitem{angerson1990lapack}
Angerson, E., Bai, Z., Dongarra, J., Greenbaum, A., McKenney, A., Du~Croz, J., Hammarling, S., Demmel, J., Bischof, C., Sorensen, D.: Lapack: A portable linear algebra library for high-performance computers. In: Supercomputing'90: Proceedings of the 1990 ACM/IEEE Conference on Supercomputing. pp. 2--11. IEEE (1990)

\bibitem{banerjee2018two}
Banerjee, A.S., Lin, L., Suryanarayana, P., Yang, C., Pask, J.E.: Two-level chebyshev filter based complementary subspace method: pushing the envelope of large-scale electronic structure calculations. Journal of chemical theory and computation  \textbf{14}(6),  2930--2946 (2018)

\bibitem{barnes2017improved}
Barnes, T.A., Kurth, T., Carrier, P., Wichmann, N., Prendergast, D., Kent, P.R., Deslippe, J.: Improved treatment of exact exchange in quantum espresso. Computer Physics Communications  \textbf{214},  52--58 (2017)

\bibitem{cances2024modified}
Canc{\`e}s, E., Hassan, M., Vidal, L.: Modified-operator method for the calculation of band diagrams of crystalline materials. Mathematics of Computation  \textbf{93}(347),  1203--1245 (2024)

\bibitem{cao2022design}
Cao, H., Chen, J.: Design and implementation of shenwei universal c/c++. arXiv preprint arXiv:2208.00607  (2022)

\bibitem{carter2008challenges}
Carter, E.A.: Challenges in modeling materials properties without experimental input. Science  \textbf{321}(5890),  800--803 (2008)

\bibitem{chen2020deepks}
Chen, Y., Zhang, L., Wang, H., E, W.: Deepks: A comprehensive data-driven approach toward chemically accurate density functional theory. Journal of Chemical Theory and Computation  \textbf{17}(1),  170--181 (2020)

\bibitem{chen2023deepks}
Chen, Y., Zhang, L., Wang, H., et~al.: Deepks-kit: A package for developing machine learning-based chemically accurate energy and density functional models. Computer Physics Communications  \textbf{282},  108520 (2023)

\bibitem{das2019fast}
Das, S., Motamarri, P., Gavini, V., Turcksin, B., Li, Y.W., Leback, B.: Fast, scalable and accurate finite-element based ab initio calculations using mixed precision computing: 46 pflops simulation of a metallic dislocation system. In: Proceedings of the International Conference for High Performance Computing, Networking, Storage and Analysis. pp. 1--11 (2019)

\bibitem{osti_1469055}
Dinge, D.: A presentation on trinity performance.  (8 2017), \url{https://www.osti.gov/biblio/1469055}

\bibitem{dongarra1990set}
Dongarra, J.J., Du~Croz, J., Hammarling, S., Duff, I.S.: A set of level 3 basic linear algebra subprograms. ACM Transactions on Mathematical Software (TOMS)  \textbf{16}(1),  1--17 (1990)

\bibitem{frigo1998fftw}
Frigo, M., Johnson, S.G.: Fftw: An adaptive software architecture for the fft. In: Proceedings of the 1998 IEEE International Conference on Acoustics, Speech and Signal Processing, ICASSP'98 (Cat. No. 98CH36181). vol.~3, pp. 1381--1384. IEEE (1998)

\bibitem{frisch1984self}
Frisch, M.J., Pople, J.A., Binkley, J.S.: Self-consistent molecular orbital methods 25. supplementary functions for gaussian basis sets. The Journal of chemical physics  \textbf{80}(7),  3265--3269 (1984)

\bibitem{gao2025pypwdft}
Gao, J., Fu, L., Jiao, S., Zhang, Z., Chen, S., Zhang, Z., Wu, W., Wan, L., Li, J., Hu, W., et~al.: Pypwdft: A lightweight python software for single-node 10k atom plane-wave density functional theory calculations. Journal of Chemical Theory and Computation  (2025)

\bibitem{giannozzi2017advanced}
Giannozzi, P., Andreussi, O., Brumme, T., Bunau, O., Nardelli, M.B., Calandra, M., Car, R., Cavazzoni, C., Ceresoli, D., Cococcioni, M., et~al.: Advanced capabilities for materials modelling with quantum espresso. Journal of physics: Condensed matter  \textbf{29}(46),  465901 (2017)

\bibitem{giannozzi2009quantum}
Giannozzi, P., Baroni, S., Bonini, N., Calandra, M., Car, R., Cavazzoni, C., Ceresoli, D., Chiarotti, G.L., Cococcioni, M., Dabo, I., et~al.: Quantum espresso: a modular and open-source software project for quantum simulations of materials. Journal of physics: Condensed matter  \textbf{21}(39),  395502 (2009)

\bibitem{groenenboom2020combined}
Groenenboom, M.C., Anderson, R.M., Wollmershauser, J.A., Horton, D.J., Policastro, S.A., Keith, J.A.: Combined neural network potential and density functional theory study of tial2o5 surface morphology and oxygen reduction reaction overpotentials. The Journal of Physical Chemistry C  \textbf{124}(28),  15171--15179 (2020)

\bibitem{gygi2006large}
Gygi, F., Draeger, E.W., Schulz, M., De~Supinski, B.R., Gunnels, J.A., Austel, V., Sexton, J.C., Franchetti, F., Kral, S., Ueberhuber, C.W., et~al.: Large-scale electronic structure calculations of high-z metals on the bluegene/l platform. In: Proceedings of the 2006 ACM/IEEE conference on Supercomputing. pp. 45--es (2006)

\bibitem{hafner2008ab}
Hafner, J.: Ab-initio simulations of materials using vasp: Density-functional theory and beyond. Journal of computational chemistry  \textbf{29}(13),  2044--2078 (2008)

\bibitem{hasegawa2011first}
Hasegawa, Y., Iwata, J.I., Tsuji, M., Takahashi, D., Oshiyama, A., Minami, K., Boku, T., Shoji, F., Uno, A., Kurokawa, M., et~al.: First-principles calculations of electron states of a silicon nanowire with 100,000 atoms on the k computer. In: Proceedings of 2011 international conference for high performance computing, networking, storage and analysis. pp. 1--11 (2011)

\bibitem{hehre1969self}
Hehre, W.J., Stewart, R.F., Pople, J.A.: Self-consistent molecular-orbital methods. i. use of gaussian expansions of slater-type atomic orbitals. The Journal of Chemical Physics  \textbf{51}(6),  2657--2664 (1969)

\bibitem{hohenberg1964inhomogeneous}
Hohenberg, P., Kohn, W.: Inhomogeneous electron gas. Phys. Rev.  \textbf{136}(3B), ~B864 (1964)

\bibitem{hu20222}
Hu, W., An, H., Guo, Z., Jiang, Q., Qin, X., Chen, J., Jia, W., Yang, C., Luo, Z., Li, J., et~al.: 2.5 million-atom ab initio electronic-structure simulation of complex metallic heterostructures with dgdft. In: 2022 SC22: International Conference for High Performance Computing, Networking, Storage and Analysis (SC). pp. 48--60. IEEE Computer Society (2022)

\bibitem{hu2017adaptively}
Hu, W., Lin, L., Banerjee, A.S., Vecharynski, E., Yang, C.: Adaptively compressed exchange operator for large-scale hybrid density functional calculations with applications to the adsorption of water on silicene. Journal of Chemical Theory and Computation  \textbf{13}(3),  1188--1198 (2017)

\bibitem{hutter2014wiley}
Hutter, J., Iannuzzi, M., Schiffmann, F., VandeVondele, J.: Wiley interdiscip. Rev.: Comput. Mol. Sci  \textbf{4}(1),  15--25 (2014)

\bibitem{2018Large}
Igram, D., Bhattarai, B., Biswas, P., Drabold, D.A.: Large and realistic models of amorphous silicon. Journal of Non-Crystalline Solids  \textbf{492},  27--32 (2018)

\bibitem{igram2018large}
Igram, D., Bhattarai, B., Biswas, P., Drabold, D.A.: Large and realistic models of amorphous silicon. Journal of Non-Crystalline Solids  \textbf{492},  27--32 (2018)

\bibitem{jensen2002polarization}
Jensen, F.: Polarization consistent basis sets. ii. estimating the kohn--sham basis set limit. The Journal of chemical physics  \textbf{116}(17),  7372--7379 (2002)

\bibitem{jia2013analysis}
Jia, W., Cao, Z., Wang, L., Fu, J., Chi, X., Gao, W., Wang, L.W.: The analysis of a plane wave pseudopotential density functional theory code on a gpu machine. Computer Physics Communications  \textbf{184}(1),  9--18 (2013)

\bibitem{jia2013fast}
Jia, W., Fu, J., Cao, Z., Wang, L., Chi, X., Gao, W., Wang, L.W.: Fast plane wave density functional theory molecular dynamics calculations on multi-gpu machines. Journal of Computational Physics  \textbf{251},  102--115 (2013)

\bibitem{jia2019parallel}
Jia, W., Wang, L.W., Lin, L.: Parallel transport time-dependent density functional theory calculations with hybrid functional on summit. In: Proceedings of the International Conference for High Performance Computing, Networking, Storage and Analysis. pp. 1--23 (2019)

\bibitem{jiang2022accelerating}
Jiang, Q., Li, J., Chen, J., Qin, X., Wan, L., Yang, J., Liu, J., Hu, W., An, H.: Accelerating parallel first-principles excited-state calculation by low-rank approximation with k-means clustering. In: Proceedings of the 51st International Conference on Parallel Processing. pp. 1--11 (2022)

\bibitem{kocabacs2023gaussian}
Kocaba{\c{s}}, T., Ke{\c{c}}eli, M., V{\'a}zquez-Mayagoitia, {\'A}., Sevik, C.: Gaussian approximation potentials for accurate thermal properties of two-dimensional materials. Nanoscale  \textbf{15}(19),  8772--8780 (2023)

\bibitem{kresse1996efficient}
Kresse, G., Furthm{\"u}ller, J.: Efficient iterative schemes for ab initio total-energy calculations using a plane-wave basis set. Physical review B  \textbf{54}(16),  11169 (1996)

\bibitem{kuhne2020cp2k}
K{\"u}hne, T.D., Iannuzzi, M., Del~Ben, M., Rybkin, V.V., Seewald, P., Stein, F., Laino, T., Khaliullin, R.Z., Sch{\"u}tt, O., Schiffmann, F., et~al.: Cp2k: An electronic structure and molecular dynamics software package-quickstep: Efficient and accurate electronic structure calculations. The Journal of Chemical Physics  \textbf{152}(19) (2020)

\bibitem{li2022deep}
Li, H., Wang, Z., Zou, N., Ye, M., Xu, R., Gong, X., Duan, W., Xu, Y.: Deep-learning density functional theory hamiltonian for efficient ab initio electronic-structure calculation. Nature Computational Science  \textbf{2}(6),  367--377 (2022)

\bibitem{lu202186}
Lu, D., Wang, H., Chen, M., Lin, L., Car, R., Jia, W., Zhang, L., et~al.: 86 pflops deep potential molecular dynamics simulation of 100 million atoms with ab initio accuracy. Computer Physics Communications  \textbf{259},  107624 (2021)

\bibitem{milman2000electronic}
Milman, V., Winkler, B., White, J., Pickard, C., Payne, M., Akhmatskaya, E., Nobes, R.: Electronic structure, properties, and phase stability of inorganic crystals: A pseudopotential plane-wave study. International Journal of Quantum Chemistry  \textbf{77}(5),  895--910 (2000)

\bibitem{motamarri2020dft}
Motamarri, P., Das, S., Rudraraju, S., Ghosh, K., Davydov, D., Gavini, V.: Dft-fe--a massively parallel adaptive finite-element code for large-scale density functional theory calculations. Computer Physics Communications  \textbf{246},  106853 (2020)

\bibitem{nakata2020large}
Nakata, A., Baker, J.S., Mujahed, S.Y., Poulton, J.T., Arapan, S., Lin, J., Raza, Z., Yadav, S., Truflandier, L., Miyazaki, T., et~al.: Large scale and linear scaling dft with the conquest code. The Journal of chemical physics  \textbf{152}(16) (2020)

\bibitem{v100}
{NVIDIA}: {NVIDIA Tesla V100 GPU Architecture. White Paper}. \url{https://images.nvidia.com/content/volta-architecture/pdf/volta-architecture-whitepaper.pdf} (2017)

\bibitem{a100}
{NVIDIA}: {NVIDIA A100 Tensor Core GPU Architecture. White Paper}. \url{https://images.nvidia.com/aem-dam/en-zz/Solutions/data-center/nvidia-ampere-architecture-whitepaper.pdf} (2020)

\bibitem{1992Iterative}
Payne, M.C., Arias, T.A., Joannopoulos, J.D.: Iterative minimization techniques for ab initio total-energy calculations: molecular dynamics and conjugate gradients. Reviews of Modern Physics; (United States)  \textbf{64:4}(4),  1045--1097 (1992)

\bibitem{pedersen2017optimal}
Pedersen, A., Pizzagalli, L., J{\'o}nsson, H.: Optimal atomic structure of amorphous silicon obtained from density functional theory calculations. New Journal of Physics  \textbf{19}(6),  063018 (2017)

\bibitem{sato2021co}
Sato, M., Kodama, Y., Tsuji, M., Odajima, T.: Co-design and system for the supercomputer “fugaku”. IEEE micro  \textbf{42}(2),  26--34 (2021)

\bibitem{shang2021extreme}
Shang, H., Li, F., Zhang, Y., Zhang, L., Fu, Y., Gao, Y., Wu, Y., Duan, X., Lin, R., Liu, X., et~al.: Extreme-scale ab initio quantum raman spectra simulations on the leadership hpc system in china. In: Proceedings of the International Conference for High Performance Computing, Networking, Storage and Analysis. pp. 1--13 (2021)

\bibitem{shang2022large}
Shang, H., Shen, L., Fan, Y., Xu, Z., Guo, C., Liu, J., Zhou, W., Ma, H., Lin, R., Yang, Y., et~al.: Large-scale simulation of quantum computational chemistry on a new sunway supercomputer. arXiv preprint arXiv:2207.03711  (2022)

\bibitem{wang2018deepmd}
Wang, H., Zhang, L., Han, J., et~al.: Deepmd-kit: A deep learning package for many-body potential energy representation and molecular dynamics. Computer Physics Communications  \textbf{228},  178--184 (2018)

\bibitem{wang2008linearly}
Wang, L.W., Lee, B., Shan, H., Zhao, Z., Meza, J., Strohmaier, E., Bailey, D.H.: Linearly scaling 3d fragment method for large-scale electronic structure calculations. In: SC'08: Proceedings of the 2008 ACM/IEEE Conference on Supercomputing. pp. 1--10. IEEE (2008)

\bibitem{wang2022scaling}
Wang, Y., Cao, H., Ma, Z., Yin, W., Chen, W.: Scaling graph 500 sssp to 140 trillion edges with over 40 million cores. In: SC22: International Conference for High Performance Computing, Networking, Storage and Analysis. pp. 1--15. IEEE (2022)

\bibitem{wang2024deeph}
Wang, Y., Li, H., Tang, Z., Tao, H., Wang, Y., Yuan, Z., Chen, Z., Duan, W., Xu, Y.: Deeph-2: enhancing deep-learning electronic structure via an equivariant local-coordinate transformer. arXiv preprint arXiv:2401.17015  (2024)

\bibitem{wells2016announcing}
Wells, J., Bland, B., Nichols, J., Hack, J., Foertter, F., Hagen, G., Maier, T., Ashfaq, M., Messer, B., Parete-Koon, S.: Announcing supercomputer summit. Tech. rep., Oak Ridge National Lab.(ORNL), Oak Ridge, TN (United States) (2016)

\bibitem{xu2023velocity}
Xu, Q., Del~Ben, M., Sait~Okyay, M., Choi, M., Ibrahim, K.Z., Wong, B.M.: Velocity-gauge real-time time-dependent density functional tight-binding for large-scale condensed matter systems. Journal of Chemical Theory and Computation  \textbf{19}(22),  7989--7997 (2023)

\bibitem{xu2021sparc}
Xu, Q., Sharma, A., Comer, B., Huang, H., Chow, E., Medford, A.J., Pask, J.E., Suryanarayana, P.: Sparc: Simulation package for ab-initio real-space calculations. SoftwareX  \textbf{15},  100709 (2021)

\bibitem{yin1982theory}
Yin, M., Cohen, M.L.: Theory of ab initio pseudopotential calculations. Physical review B  \textbf{25}(12), ~7403 (1982)

\bibitem{zeng2023deepmd}
Zeng, J., Zhang, D., Lu, D., Mo, P., Li, Z., Chen, Y., Rynik, M., Huang, L., Li, Z., Shi, S., et~al.: Deepmd-kit v2: A software package for deep potential models. The Journal of Chemical Physics  \textbf{159}(5) (2023)

\bibitem{zeni2025generative}
Zeni, C., Pinsler, R., Z{\"u}gner, D., Fowler, A., Horton, M., Fu, X., Wang, Z., Shysheya, A., Crabb{\'e}, J., Ueda, S., et~al.: A generative model for inorganic materials design. Nature pp.~1--3 (2025)

\bibitem{zhang2025advancing}
Zhang, C., Zhong, Y., Tao, Z.G., Qin, X., Shang, H., Lan, Z., Prezhdo, O.V., Gong, X.G., Chu, W., Xiang, H.: Advancing nonadiabatic molecular dynamics simulations in solids with e (3) equivariant deep neural hamiltonians. Nature Communications  \textbf{16}(1), ~2033 (2025)

\bibitem{zhang2022uniq}
Zhang, C., Wang, H., Ma, Z., Xie, L., Song, Z., Zhai, J.: Uniq: a unified programming model for efficient quantum circuit simulation. In: 2022 SC22: International Conference for High Performance Computing, Networking, Storage and Analysis (SC). pp. 692--707. IEEE Computer Society (2022)

\bibitem{zhao2018porting}
Zhao, H., Zhang, C.: Porting and optimizing vasp on the sw26010. In: Algorithms and Architectures for Parallel Processing: ICA3PP 2018 International Workshops, Guangzhou, China, November 15-17, 2018, Proceedings. vol. 11338, p.~17. Springer (2018)

\bibitem{zhao2022ai}
Zhao, X., Li, M., Xiao, Q., Chen, J., Wang, F., Shen, L., Zhao, M., Wu, W., An, H., He, L., et~al.: Ai for quantum mechanics: high performance quantum many-body simulations via deep learning. In: 2022 SC22: International Conference for High Performance Computing, Networking, Storage and Analysis (SC). pp. 677--691. IEEE Computer Society (2022)

\bibitem{zhao2017performance}
Zhao, Z., Marsman, M., Wende, F., Kim, J.: Performance of hybrid mpi/openmp vasp on cray xc40 based on intel knights landing many integrated core architecture. CUG Proceedings  (2017)

\bibitem{0Enabling}
Zhu, Q., Luo, H., Yang, C., Ding, M., Yin, W., Yuan, X.: Enabling and scaling the hpcg benchmark on the newest generation sunway supercomputer with 42 million heterogeneous cores. In: SC21: International Conference for High Performance Computing, Networking, Storage and Analysis

\end{thebibliography}

\end{document}